\documentclass[multphys,vecphys]{svmult}

\usepackage{makeidx}         
\usepackage{graphicx}        
\usepackage{multicol}        

\usepackage[bottom]{footmisc}

\makeindex             

\begin{document}

\title*{Half-metallicity and Slater-Pauling behavior in the 
ferromagnetic Heusler alloys}

\titlerunning{Slater-Pauling behavior and half-metallicity in Heusler alloys}

\author{Iosif Galanakis and
Peter H. Dederichs}

\institute{ Institut f\"ur
Festk\"orperforschung, Forschungszentrum J\"ulich, D-52425
J\"ulich, Germany \texttt{I.Galanakis@fz-juelich.de and P.H.Dederichs@fz-juelich.de}}

\maketitle

\section{Introduction}
\label{secios:1}

\begin{figure}
\centering
\includegraphics[scale=0.4]{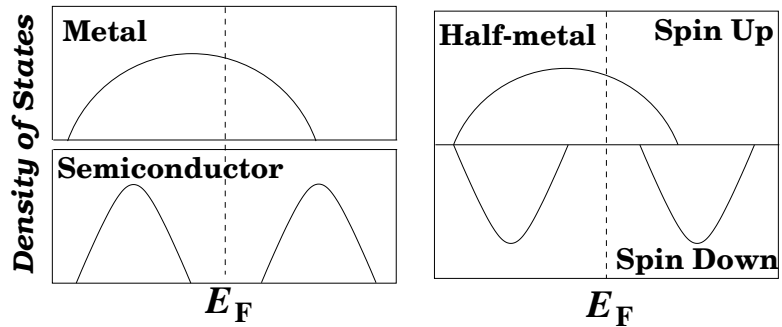}
\caption{Schematic representation of the density of states 
for a half-metal with respect to normal metals and 
semiconductors.} \label{figios1a}
\end{figure}

Half-metallic ferromagnets represent a new class of materials which attracted a lot of
attention due to their possible applications in
spintronics (also known as magnetoelectronics) \cite{Zutic2004}.  
Adding the spin degree of freedom to the
conventional electronic devices has several advantages like 
non-volatility, increased data processing speed, decreased
electric power consumption and  increased integration
densities \cite{Wolf}. The current advances in new materials 
and especially in the half-metals are
promising for engineering new spintronic devices in the near
future \cite{Wolf}.  In these materials the two spin bands
show a completely different behavior. While the majority spin band
(referred also as spin-up band) shows the typical metallic
behavior, the minority spin band (spin-down band) exhibits a 
semiconducting behavior with a gap at the Fermi level. 
Therefore such half-metals are ferromagnets and can be considered
as hybrids between metals and semiconductors. A schematic representation of the 
density of states of a half-metal as compared to a normal metal 
and a normal semiconductor is shown in figure \ref{figios1a}. 
The spinpolarization at the Fermi level is 100\%
and therefore these compounds should have a fully spinpolarised 
current and might be able to yield a 100\% spininjection and 
thus to maximize the efficiency of
magnetoelectronic devices \cite{deBoeck2002}.

Heusler alloys \cite{heusler} have attracted during the last
century a great interest due to the possibility to study in the
same family of alloys a series of interesting diverse magnetic
phenomena like itinerant and localized magnetism,
antiferromagnetism, helimagnetism, Pauli paramagnetism or
heavy-fermionic behavior \cite{landolt,landolt2,Pierre97,Tobola}. 
The first Heusler alloys studied were crystallizing in the $L2_1$
structure which consists of 4 fcc sublattices. Afterwards, it was
discovered that it is possible to leave one of the four
sublattices unoccupied ($C1_b$ structure). The latter
compounds are often called half- or semi-Heusler alloys, while the $L2_1$
compounds are referred to as full-Heusler alloys. In a pioneering theory 
paper in 1983 de Groot and  his collaborators \cite{groot} showed 
by using first-principles electronic structure calculations that one of the
half-Heusler compounds, NiMnSb, is a half-metal, i.e. the
minority band is semiconducting with a gap at the Fermi level
$E_F$, leading to 100\% spin polarization at $E_F$ as shown in figure
\ref{figios1a}. 
Other known half-metallic materials except the 
half- and full-Heusler alloys 
\cite{Galanakis2002a,Galanakis2002b,GalanakisQuart,Zhang} 
are some oxides (\textit{e.g} CrO$_2$ and Fe$_3$O$_4$) \cite{Soulen98}, the manganites
(\textit{e.g} La$_{0.7}$Sr$_{0.3}$MnO$_3$) \cite{Soulen98},  
the double perovskites (\textit{e.g.} Sr$_2$FeReO$_6$) \cite{Kato},
the pyrites (\textit{e.g} CoS$_2$) \cite{Pyrites}, the
transition metal chalcogenides (\textit{e.g} CrSe) and pnictides 
(\textit{e.g} CrAs) in the
zinc-blende or wurtzite structures \cite{GalanakisZB,Xie,Akinaga,Zhao},
the europium chalcogenides (\textit{e.g} EuS) \cite{Temmerman} and the diluted
magnetic semiconductors (\textit{e.g} Mn impurities in Si or
GaAs)\cite{FreemanMnSi,Akai98}.
 Although thin films of CrO$_2$
and La$_{0.7}$Sr$_{0.3}$MnO$_3$ have been verified to present
practically 100\% spin-polarization at the Fermi level at low
temperatures \cite{Soulen98,Park98}, the Heusler alloys remain attractive for
technical applications like spin-injection devices \cite{Data},
spin-filters \cite{Kilian00}, tunnel junctions \cite{Tanaka99}, or
GMR devices \cite{Caballero98} due to their  relatively  high  Curie
temperature compared to these compounds \cite{landolt}.

The half-metallic character of NiMnSb in single crystals seems to
have been well-established experimentally. Infrared
absorption \cite{Kirillova95} and spin-polarized
positron-annihilation \cite{Hanssen90} gave a spin-polarization of
$\sim$100\% at the Fermi level. Recently it has also become possible 
to grow high quality films of
Heusler alloys, and it is mainly NiMnSb that has attracted the
attention \cite{Roy,Bach,Giapintzakis}. Unfortunately these films were found not to
be half-metallic      \cite{Soulen98,Mancoff99,Zhu01,Bona,Clowes}; 
a maximum value of 58\% for the spin-polarization of NiMnSb was obtained by
Soulen \textit{et al.} \cite{Soulen98}. These polarization values
are consistent with a small perpendicular magnetoresistance
measured for NiMnSb in a spin-valve structure \cite{Caballero99},
a superconducting tunnel junction \cite{Tanaka99} and a tunnel
magnetoresistive junction \cite{Tanaka97}. 
Ristoiu \textit{et al.}
showed that during the growth of the NiMnSb thin films, Sb and
then Mn atoms segregate to the surface, which is far from being
perfect, thus decreasing the obtained spin-polarization
\cite{Ristoiu00}. But when they removed the excess of Sb by
flash annealing, they managed to get a nearly stoichiometric
ordered alloy surface being terminated by a MnSb layer, which
presented a spin-polarization of about 67$\pm$9\% at room
temperature \cite{Ristoiu00}. 

Several groups have verified the half-metallic character of
bulk NiMnSb using first-principles
calculations \cite{iosif,Calculations}. Larson \textit{et al.} have shown
that the actual structure of NiMnSb is the most stable with
respect to an interchange of the atoms \cite{Larson00} and Orgassa
\textit{et al.} showed that a few percent of disorder induce
states within the gap but do not destroy the
half-metallicity \cite{Orgassa99}. Recently, Galanakis has 
shown by first-principle
calculations that NiMnSb surfaces do not present 100\%
spin-polarization \cite{GalanakisSurf} but  Wijs and de Groot 
proposed that at some interfaces it is
possible to restore the half-metallic character of NiMnSb
\cite{groot2}. These results were also
confirmed by Debernardi \textit{et al.} who studied the interface
between NiMnSb and GaAs \cite{Debernardi}. Jenkins and King studied by a
pseudopotential technique the MnSb terminated (001) surface of
NiMnSb and showed that there are two surface states at the Fermi
level, which are well localized at the surface layer
\cite{Jenkins01} and they persist even when the MnSb surface is covered by a
Sb overlayer \cite{Jenkins02}. 

\begin{figure}
\centering
\includegraphics[scale=0.4]{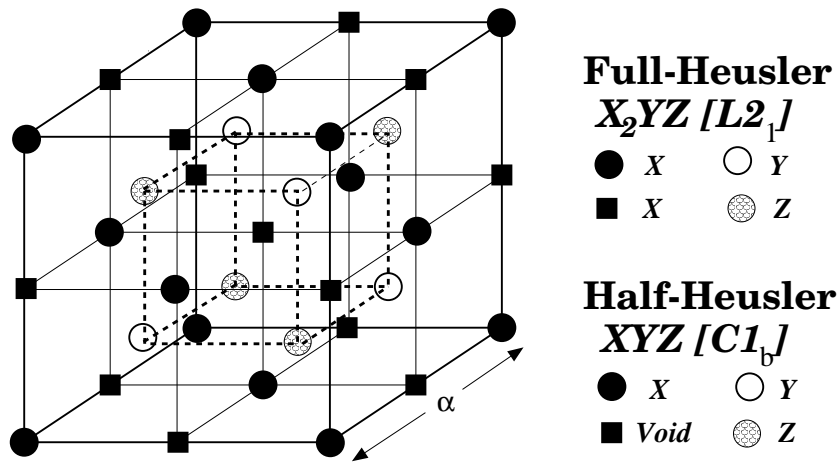}
\caption{$C1_b$ and $L2_1$ structures adapted by the half- and full-Heusler alloys.
The lattice is consisted of 4 interprenatating f.c.c. lattice. 
The unit cell is that of a fcc lattice with four
atoms as basis, \textit{e.g.} CoMnSb: Co at $(0\:0\:0)$, Mn at
$({1\over4}\:{1\over4}\:{1\over4})$, a vacant site at
$({1\over2}\:{1\over2}\:{1\over2})$  and Sb at
$({3\over4}\:{3\over4}\:{3\over4})$  in Wyckoff coordinates. In the case of the full
Heusler alloys also the vacant site is occupied by a Co atom.
Note also that if all atoms were identical, the lattice would be simply the 
bcc.} \label{figios1}
\end{figure}

Webster and Ziebeck  \cite{Webster} and Suits \cite{Suits76}  were the 
first to synthesize full-Heusler alloys  containing Co and Rh,
respectively. K\"ubler \textit{et al.} studied the mechanisms stabilizing 
the ferro- or the antiferromagnetism in these compounds \cite{Kubler83}. Ishida
and collaborators  have proposed that the
compounds of the type Co$_2$MnZ, where Z stands for Si and Ge, are
 half-metals \cite{Ishida95,Fujii90}. Also the Heusler alloys of the type
Fe$_2$MnZ have been proposed to show
half-metallicity \cite{Fujii95}. But Brown \textit{et
al.} \cite{Brown00}  using polarized neutron diffraction
measurements have shown that there is a finite very small
spin-down density of states (DOS) at the Fermi level instead of an
absolute gap in agreement with the \textit{ab-initio} calculations
of K\"ubler \textit{et al.} for the Co$_2$MnAl and Co$_2$MnSn
compounds \cite{Kubler83}.  Recently, several groups 
managed to grow  Co$_2$MnGe and  Co$_2$MnSi thin films on various substrates 
 \cite{Ambrose,Raphael,Chen2004}, and there also exist
first-principles calculations for the (001) surface of such an
alloy \cite{Ishida98}. Geiersbach and collaborators
have grown (110) thin films of Co$_2$MnSi, Co$_2$MnGe and
Co$_2$MnSn using a metallic seed on top of a MgO(001)
substrate \cite{Geiersbach} and studied also the transport properties
of multilayers of these compounds with normal metals \cite{Westerholt}.
But as Picozzi \textit{et al.} have shown the interfaces of such
structures are not half-metallic \cite{picozziMult}. Finally, K\"ammerer 
and collaborators managed to built magnetic tunnel junctions based on
Co$_2$MnSi and found a tunneling magnetoresistance effect much larger 
than when the Ni$_{0.8}$Fe$_{0.2}$ or Co$_{0.3}$Fe$_{0.7}$ are 
used as magnetic electrodes \cite{Kammerer}. Similar experiments 
have been undertaken by Inomata and collaborators using 
Co$_2$Cr$_{0.6}$Fe$_{0.4}$Al as the magnetic electrode \cite{Inomata}.

In this contribution,  we  present a study of the basic electronic and magnetic 
properties of the half-metallic Heusler alloys. 
Analyzing the  \textit{ab-initio} results using the 
group-theory and simple models we explain the origin of the gap in both the half- and 
full-Heusler alloys, which is fundamental for 
understanding their electronic and magnetic properties.
For both families of compounds the total 
spin magnetic moment scales with the number of valence electron, thus opening the way 
to engineer new half-metallic Heusler alloys with the desired magnetic properties.
Although in general the surfaces loose the half-metallic 
character and show only a small degree of spin-polarization, we show that in the 
case of compounds containing Cr, the very large Cr moments at the surface reduce the
importance of the surface states and the spin-polarization of such surfaces is very high,
\textit{e.g} 84\% for the CrAl-terminated Co$_2$CrAl(001) surface. 
surface. Finally we discuss the role of  defects and spin-orbit coupling on the 
half-metallic band gap.

In sections \ref{secios:2} and \ref{secios:3} 
we present the electronic and magnetic properties  of the XMnSb
(X=Ni,Co,Rh,Pd,Ir or Pt) and Co$_2$MnZ (Z=Al,Si,Ga,Ge or Sn) compounds, respectively.
In section \ref{secios:4} we investigate the effect of compressing or 
expanding  the 
lattice and in
section \ref{secios:5} the properties of the quaternary Heusler alloys.
In sections \ref{secios:6} and \ref{secios:7} we review the role of defects and 
of the spin-orbit coupling, respectively and finally in section \ref{secios:8}
we review the surface properties of these alloys.

\section{Electronic Structure and Magnetism of Half-Heusler Alloys}
\label{secios:2}

\subsection{Band Structure of Half-Heusler Alloys}
\label{secios:2-1}

\begin{figure}
\centering
\includegraphics[scale=0.6]{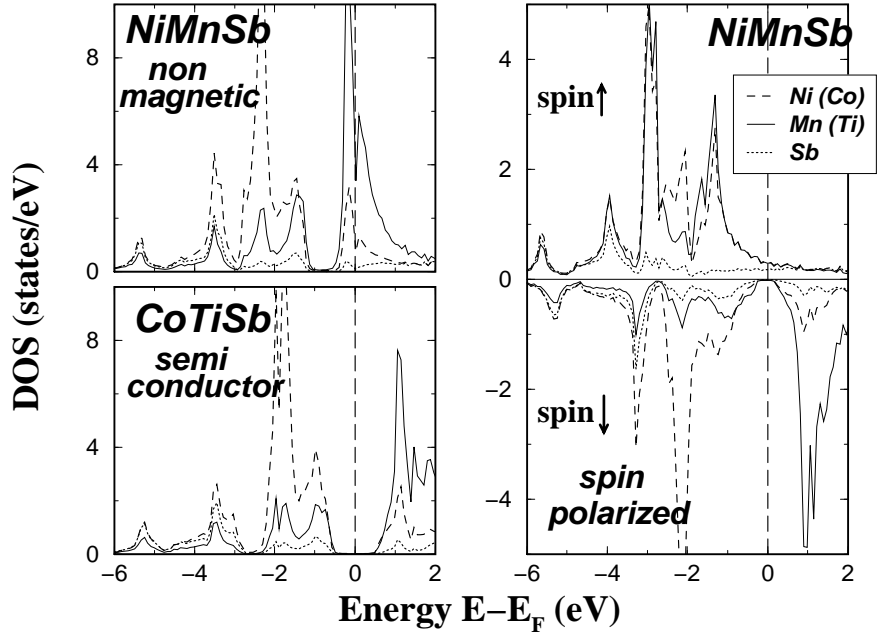}
\caption{Atom-resolved density of states (DOS) of NiMnSb for a paramagnetic (left) and
ferromagnetic (right) calculation. The zero energy value corresponds to the Fermi
level $E_\mathrm{F}$} \label{figios2}
\end{figure}

In the following we present results for the densities of states of some typical 
half-Heusler alloys of C1$_b$ structure (see figure \ref{figios1}), 
sometimes also referred to as semi-Heusler 
alloys. To perform the calculations, we used the Vosko, Wilk and Nusair
parameterization \cite{Vosko} for the local density approximation
(LDA) to the exchange-correlation potential \cite{kohn} to solve
the Kohn-Sham equations within the full-potential screened
Korringa-Kohn-Rostoker (FSKKR) method \cite{Zeller95,Papanikolaou02}.   
The prototype example is NiMnSb, the half-metal discovered in 1983 by de Groot 
\cite{groot}. Figure \ref{figios2} shows the density of states (DOS) of NiMnSb in 
a non-spin-polarized calculation (left upper panel) and in a calculation correctly 
including the spin-polarization (right panel). Given are the local 
contributions to the density of states 
(LDOS) on the Ni-site (dashed), the Mn-site (full line) and the Sb-site (dotted). 
In the non-magnetic case the DOS of NiMnSb has contributions from 4 different bands: 
Each Sb atom with the atomic configuration 5$s^2$5$p^3$ introduces a deep lying $s$ band, which is located at about -12eV and is not shown in the figure, and three $p$-bands 
in the regions between -5.5 and -3eV. These bands are separated by a deep minimum in 
the DOS from 5 Ni $d$ bands between -3 and -1eV, which themselves are separated by a 
sizeable band gap from the upper 5 $d$-bands of Mn. Since all atomic orbitals, i.e. the 
Ni $d$, the Mn $d$  and the Sb $sp$ orbitals hybridize with each other, all bands are 
hybrids between these states, being either of bonding or antibonding type. Thus the Ni 
$d$-bands contain a bonding Mn $d$ admixture, while the higher Mn $d$-bands are 
antibonding hybrids with small Ni $d$-admixtures. Equally the Sb $p$-bands exhibit 
strong Ni $d$- and somewhat smaller Mn $d$-contributions.  

This configuration for NiMnSb is energetically not stable, since (i) the Fermi energy 
lies in the middle of an antibonding band and (ii) since the Mn atom can gain 
considerable exchange energy by forming a magnetic moment. Therefore the spin-polarized 
results (right figure) show a considerably different picture. In the majority 
(spin $\uparrow$) band the Mn $d$ states are shifted to lower energies and form a common 
$d$ band with the Ni $d$ states, while in the minority band (spin $\downarrow$) the 
Mn states are shifted to higher energies and are unoccupied, so that a band gap at 
$E_F$ is formed separating the occupied $d$ bonding from the unoccupied $d$-type 
antibonding states. Thus NiMnSb is a half-metal, with 
a band gap at $E_F$ in the minority band and a metallic DOS at $E_F$ in the majority 
band. The total magnetic moment, located mostly at the Mn atom, can be easily estimated 
to be exactly 4 $\mu_B$. Note that NiMnSb has 22 valence electrons per unit cell, 10 
from Ni, 7 from Mn and 5 from Sb. 
Since, due to the gap at $E_F$, in the minority band exactly 9 bands are fully occupied
(1 Sb-like $s$ band, 3 Sb-like p bands and 5 Ni-like $d$ bands) 
and  accommodate 9 electrons
per unit cell, the majority band contains 22 - 9 = 13 electrons, resulting in a moment 
of 4 $\mu_B$ per unit cell. 

The above non-spinpolarized calculation for NiMnSb (figure \ref{figios2} left
upper panel) suggests, 
that if we could shift the Fermi energy as in a rigid band model, a particular stable 
compound would be obtained if the Fermi level falls for both spin directions into the 
band gap. Then for both spin directions 9 bands would be occupied, resulting in a 
semiconductor with 18 valence electrons. Such semiconducting Heusler alloys indeed 
exist. As a typical example, Figure \ref{figios2} shows also the DOS of 
CoTiSb, which has 
a gap of 0.8 eV \cite{Tobola}. The gap is indirect corresponding to transitions from 
the valence band maximum at $\Gamma$ to the conduction band minimum at $X$. Other such 
semiconductors are CoZrSb (0.8 eV), FeVSb (0.36 eV) and NiTiSn (0.14 eV) where
the values in the bracket denote the size of the gap \cite{Tobola}.

\subsection{XMnSb Half-Heusler Alloys with X = Ni, Pd Pt and Co, Rh, Ir}
\label{secios:2-2}

\begin{figure}
\centering
\includegraphics[scale=0.6]{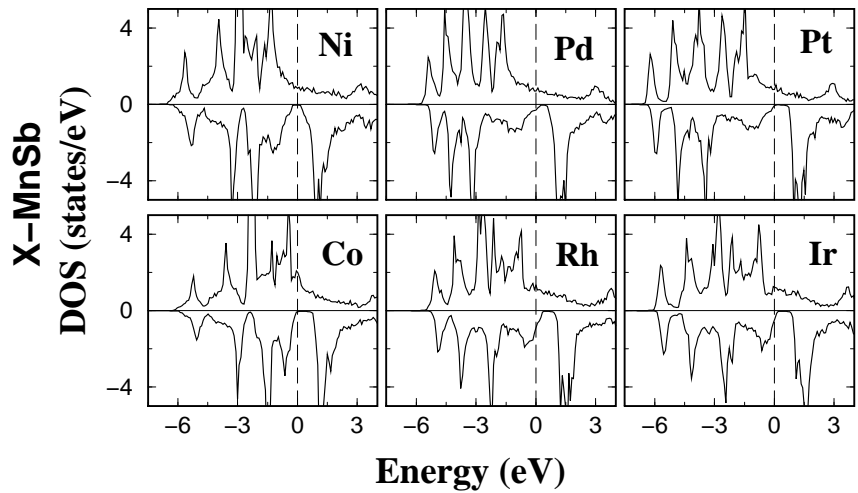}
\caption{DOS of XMnSb compounds for X= Ni, Pd, Pt and Co, Rh, Pd.} 
\label{figios3}
\end{figure}

Here we present the electronic structure of the half-Heusler
alloys of the type XMnSb, with X being an element of the Co or Ni
columns in the periodic table. These compounds are known
experimentally to be ferromagnets with high Curie temperatures
ranging between 500 K and 700 K for the Co, Ni, Pd and Pt
compounds, while the Curie temperatures of the Ir and Rh compounds
are around room temperature \cite{landolt}. In figure \ref{figios3} we
present the spin-projected total density of states (DOS) for  all
the six compounds. We remark that all compounds present a gap,
which is wider in the compounds containing Co, Rh or Ir than in
Ni, Pd or Pt. As above Sb $p$ states occupy the lowest part of the DOS
shown in the figure, while the Sb $s$ states are located $\sim$12
eV below the Fermi level. For the Ni compound the Fermi level is
at the middle of the gap and for PtMnSb at the left edge of the
gap in agreement with previous FPLMTO calculations 
\cite{iosif}.  In the case of
CoMnSb the gap is considerably larger ($\sim$1 eV) than in the
previous two compounds and the Fermi  level is located at the left
edge of the spin-down gap. CoMnSb has been studied previously
by K\"ubler, who found similar results by using the ASW method.
For the other three compounds
the Fermi level is located below the gap, although in the case of
PdMnSb and IrMnSb it is close to the band edge.

The DOS of the different systems are mainly characterized by the
large exchange-splitting of the Mn $d$ states which is around 3 eV
in all cases. This 
leads to large localized spin moments at the Mn site, the
existence of which has been verified also
experimentally \cite{Yablonskikh01}. The localization comes from
the fact that although $d$ electrons of Mn are itinerant, the
spin-down electrons are almost excluded from the Mn site. In table
\ref{tableios:1} we present the spin magnetic moments at the
different sites for all the compounds under study. The moments are calculated by integrating the
spin-projected charge density  inside every Wigner-Seitz
polyhedron. Experimental
values for the spin-moment at the Mn site can be deduced from the
experiments of Kimura \textit{et al.} \cite{Kimura97} by applying
the sum rules to their x-ray magnetic circular dichroism spectra
and the extracted moments agree nicely with our results; they found
a Mn spin moment of 3.85 $\mu_B$ for NiMnSb, 3.95 $\mu_B$ for
PdMnSb and 4.02 $\mu_B$ for PtMnSb. In the case of the Co-, Rh-,
and IrMnSb compounds the spin magnetic moment of the X atom is
antiparallel to the Mn localized moment and the Mn moment is
generally about 0.5 $\mu_B$ smaller than in the Ni, Pd and Pt
compounds. The Sb atom is here again antiferromagnetically coupled
to the Mn atom.

\begin{table}
\centering \caption{Calculated spin magnetic moments in $\mu_B$
for the XMnSb compounds. (The experimental lattice constants \protect{\cite{landolt}} 
have been used.)}
\label{tableios:1}
\begin{tabular}{r|r|r|r|r|r}
\hline\noalign{\smallskip}
 $m^{spin}(\mu_B)$ & X & Mn & Sb & Void
& Total\\ \noalign{\smallskip}\hline\noalign{\smallskip}
 NiMnSb & 0.264 & 3.705 & -0.060 & 0.052 & 3.960\\
PdMnSb & 0.080 & 4.010 & -0.110 & 0.037 &4.017 \\ PtMnSb & 0.092 &
3.889 & -0.081 &
0.039 &3.938\\ CoMnSb & -0.132 & 3.176 & -0.098 & 0.011& 2.956 \\
RhMnSb & -0.134 & 3.565 & -0.144 & $<$0.001 & 3.287 \\ IrMnSb &
-0.192 & 3.332 & -0.114 & -0.003 & 3.022 \\ FeMnSb & -0.702 &
2.715 & -0.053 & 0.019  & 1.979 \\ \noalign{\smallskip}\hline
\end{tabular}
\end{table}

The total  magnetic moment in $\mu_B$  is just the difference
between the number of spin-up occupied states and the spin-down
occupied states. 
As explained above, the number of occupied spin-down states is given by the number 
of spin down bands, i.e. 9, so that the number of occupied spin-up states is 22-9 = 13 
for NiMnSb and the isovalent compounds with Pd and Pt,
but 21-9 = 12 for CoMnSb, RhMnSb and IrMnSb and 20-9 = 11 for FeMnSb, provided
that the Fermi level stays within the gap.
Therefore one expects  total moments of 4$\mu_B$ for Ni-, Pd- and PtMnSb, 3 $\mu_B$ for 
the compounds with Co, Rh and Ir and 2 $\mu_B$ for FeMnSb. In general, for a total 
number $Z_t$ of valence electrons in the unit cell, the total moment $M_t$ is given 
by $M_t = Z_t - 18$, since with 9 electron states occupied in the minority band, 
$Z_t - 18$ is just the number of uncompensated electron spins.  

The local moment per unit
cell as given in table \ref{tableios:1} is close to 4 $\mu_B$ in
the case of NiMnSb, PdMnSb and PtMnSb, which is in agreement with
the half-metallic character (or nearly half-metallic
character in the case of PdMnSb) observed in figure
\ref{figios3}. Note that due to problems with the $\ell_{max}$
cutoff the KKR method can only give the correct integer number 4,
if Lloyd's formula has been used in the evaluation of the
integrated density of states, which is not the case in the present
calculations. We also find that the local moment of Mn is not far
away from the total number of 4 $\mu_B$ although there are significant (positive)
contributions from the X-atoms and a negative contribution from
the Sb atom. In contrast to this we find that for the
half-metallic CoMnSb and IrMnSb compounds the total moment is
about 3 $\mu_B$. Also the local moment of Mn is reduced, but only
by about 0.5 $\mu_B$. The reduction of the total moment to 3
$\mu_B$ is therefore accompanied by negative Co and Ir spin
moments, \textit{i.e} these atoms couple antiferromagnetically to
the Mn moments.  The hybridization between Co and Mn is
considerably larger than between Ni and Mn being a consequence of the smaller electronegativity difference and the larger extend of the Co orbitals. Therefore the minority
valence band of CoMnSb has  a larger Mn admixture than the one of
NiMnSb whereas the minority conduction band of CoMnSb has a larger
Co admixture than the Ni admixture in the NiMnSb conduction band,
while the populations of the majority bands are barely changed. As
a consequence, the Mn moment is reduced by the increasing
hybridization, while the Co moment becomes negative, resulting
finally in a reduction of the total moment from 4 to 3 $\mu_B$.
The table also shows that further substitution of Fe for Co
leads also to a half-metallic alloy with a total spin
magnetic moment of 2 $\mu_B$ as has been already shown by de Groot
\textit{et al.} in reference~\cite{Groot86}.

\subsection{Origin of the Gap}
\label{secios:2-3}

\begin{figure}
\centering
\includegraphics[height=5cm]{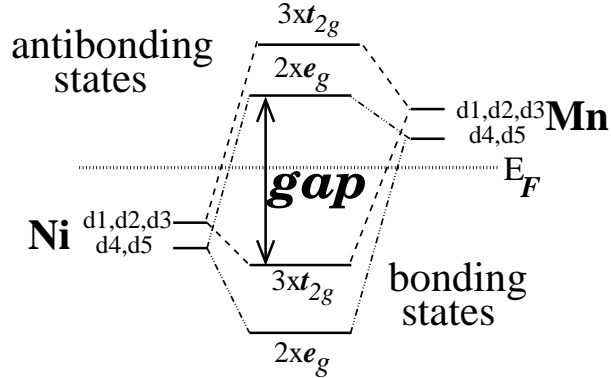}
\caption{Schematic illustration of the origin of the gap in the minority band in 
half-Heusler alloys and in compound semiconductors: The energy levels $E_b$ of the 
energetically lower lying bonding hybrids are separated from the levels $E_{ab}$ of 
the antibonding hybrids by a gap, such that only the bonding states are occupied.} 
\label{figios4}
\end{figure}

The inspection of the local DOS shown in figure \ref{figios2} for the ferromagnet 
NiMnSb as well as for the semiconductor CoTiSb shows that the DOS close to the gap is 
dominated by $d$-states: in the valence band by bonding hybrids with large Ni or Co 
admixture and in the conduction band by the antibonding hybrids with large Mn or Ti 
admixture. Thus the gap originates from the strong hybridization 
between the $d$ states of the higher valent and the lower valent transition 
metal atoms. This is shown schematically in figure \ref{figios4}. Therefore the 
origin of the gap is very similar to the gap in compound semiconductors like 
GaAs which is enforced by the hybridization of the lower lying As $sp$-states 
with the energetically higher Ga $sp$-states. 
Note that in the C1$_b$-structure the Ni and Mn sublattices form a zinc-blende 
structure, which is important for the formation the gap. The difference with 
respect to GaAs is than only, that 5 $d$-orbitals, i.e. 3 $t_{2g}$ and 3 $e_g$ orbitals, 
are involved in the hybridization, instead of 4 $sp^3$-hybrids in the compound 
semiconductors. 

\begin{figure}
\centering
\includegraphics[scale=0.6]{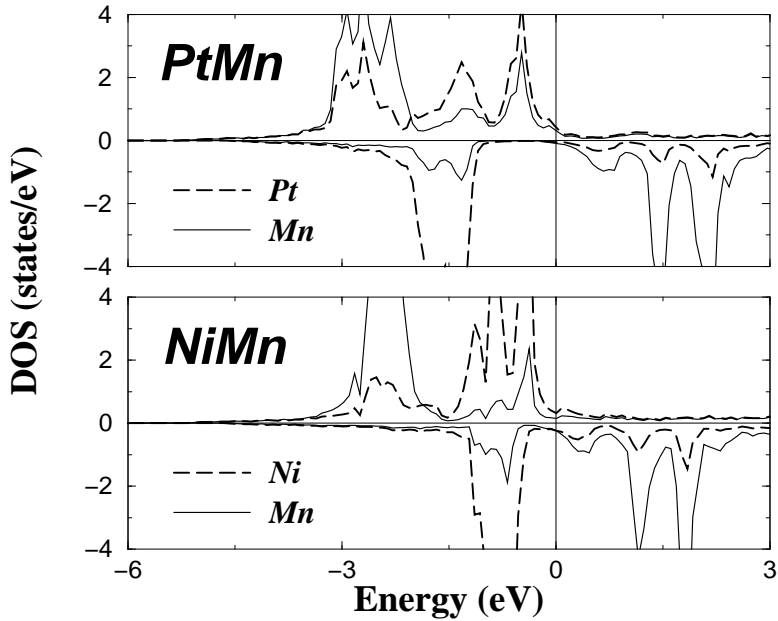}
\caption{Atom-resolved DOS for the hypothetical PtMn and NiMn crystallizing in
the zinc-blende structure.} \label{figios5}
\end{figure}

Giving these arguments it is tempting to claim, that also a hypothetical zinc-blende 
compound like NiMn or PtMn should show a half-metallic character with a gap at $E_F$ 
in the minority band. Figure \ref{figios5} shows the results of a 
self-consistent calculation for 
such zinc-blende NiMn and PtMn, with the same lattice constant as NiMnSb. Indeed a gap 
is formed in the minority band. In the hypothetical NiMn the Fermi energy is slightly 
above the gap, however the isoelectronic PtMn compound shows indeed half-metallicity. 
In this case the occupied minority bands consists of six bands,
a low-lying $s$-band and five 
bonding $d$-bands, both of mostly Pt character. 
Since the total number of valence electrons 
is 17, the majority bands contain 11 electrons, so that the total moment per unit 
cell is $11\:-\: 6\:=\: 5 \mu_B$, which is indeed obtained in 
the calculations. This is the largest 
possible moment for this compound, since in the minority band all 5 Mn $d$-states 
are empty while all majority $d$-states are occupied. The same limit of $5 \mu_B$ 
is also the maximal possible moment of the half-metallic $C1_b$ Heusler alloys. 

The gap in the half-metallic $C1_b$ compounds is normally an indirect gap, with the 
maximum of the valence band at the $\Gamma$ point and the minimum of the conduction 
band at the $X$-point. For NiMnSb we obtain a band gap of about 0.5 eV, which is in 
good agreement with the experiments of Kirillova and collaborators \cite{Kirillova95}, who, 
analyzing their infrared spectra, estimated a gap width of $\sim 0.4$ eV. As seen 
already from figure \ref{figios3} the gap of CoMnSb is considerable larger 
($\sim 1$ eV) and the  Fermi level is located at the edge of the minority valence band.

As it is well-known, the local density approximation (LDA) and the generalized 
gradient approximation (GGA) strongly underestimate the values of the gaps in 
semiconductors, typically by a factor of two. However, very good values for these 
gaps are obtained in the so-called GW approximation of Hedin and Lundqvist \cite{Hedin}, which 
describes potential in semiconductors very well. On the other hand the minority 
gap in the half-metallic systems might be better described by the LDA and 
GGA since in these system the screening is metallic. 

\subsection{Role of $sp$-Elements}
\label{secios:2-4}

While the $sp$-elements are not responsible for the existence of the minority gap, 
they are nevertheless very important for the physical properties of the Heusler alloys 
and the structural stability of the $C1_b$ structure, as we discuss in the following. 

While an Sb atom has 5 valence electrons (5$s^2$, 5$p^3$), in the
NiMnSb compound each Sb atom introduces a deep lying $s$-band, at
about -12 eV, and three $p$-bands below the center of the
$d$-bands. These bands accommodate a total of 8 electrons per unit
cell, so that formally Sb acts as a triple charged Sb$^{-3}$ ion.
Analogously, a Te-atom behaves in these compounds as a Te$^{-2}$
ion and a Sn-atom as a Sn$^{-4}$ ion. This does not mean, that
locally such a large charge transfer exists. In fact, the $s$- and
$p$-states strongly hybridize with the TM $d$-states and the
charge in these bands is delocalized and
locally Sb even looses about one electron, if one counts the charge in the 
Wigner-Seitz cells. What counts is
that the $s$- and $p$-bands accommodate 8 electrons per unit cell,
thus effectively reducing the $d$-charge of the TM atoms. 

This is nicely illustrated by the existence of the semiconducting compounds 
CoTiSb and NiTiSn. Compared to CoTiSb, in NiTiSn the missing $p$-charge of 
the Sn atom is replace by an increased $d$ charge of the Ni atom, so that 
in both cases all 9 valence bands are occupied.

The $sp$-atom is very important for the structural stability of the Heusler alloys. 
For instance, it is difficult to imagine that 
the calculated half-metallic NiMn and PtMn 
alloys with zinc-blende structure, the LDOS of which are shown in figure \ref{figios5}, 
actually  exist, since metallic alloys prefer high coordinated structures like fcc, 
bcc, hcp etc. Therefore the $sp$-elements are decisive of the 
stability of the $C1_b$ compounds. A careful discussion of the 
bonding in these compounds has been recently published by Nanda and Dasgupta 
\cite{nanda-dasgupta} using the crystal orbital Hamiltonian population (COHP) method. 
For the semiconductor FeVSb they find that while the largest contribution to the 
bonding arises from the V-$d$ -- Fe-$d$ hybridization, contributions of similar size 
arise also from the Fe-$d$ -- Sb-$p$ and the V-$d$ -- Sb-$p$ hybridization. 
Similar results are also valid for the semiconductors like CoTiSb and NiTiSn and 
in particular for the half-metal NiMnSb. Since the majority $d$-band is completely 
filled, the major part of the bonding arises from the minority band, so that 
similar arguments as for the semiconductors apply. 

Another property of the $sp$-elements is worthwhile to mention: substituting the 
Sb atom in NiMnSb by Sn, In or Te destroys the half-metallicity \cite{Galanakis2002a}. This 
is in contrast to the substitution of Ni by Co or Fe, which is documented in table 
\ref{tableios:1}. The total moment of 4 $\mu_B$ for NiMnSb is reduced to 3 $\mu_B$ in 
CoMnSb and 2 $\mu_B$ in FeMnSb, thus preserving half-metallicity. In NiMnSn the total 
moment is reduced to 3.3 $\mu_B$ (instead of 3) and in NiMnTe the total moment 
increases only to 4.7 $\mu_B$ (instead of 5). Thus by changing the $sp$-element it 
is rather difficult to preserve the half-metallicity, since the density of states 
changes more like in a rigid band model \cite{Galanakis2002a}.

\subsection{Slater-Pauling Behavior}
\label{secios:2-5}

As discussed above the total moment of the half-metallic $C1_b$ Heusler alloys 
follows the simple rule: $M_t = Z_t - 18$, where $Z_t$ is the total number of 
valence electrons. In short, the total number of electrons $Z_t$ is given by 
the sum of the number of spin-up and spin-down electrons, while the total moment 
$M_t$ is given by the difference
\begin{equation}
Z_t = N_\uparrow + N_\downarrow \quad , \quad
M_t = N_\uparrow - N_\downarrow \quad \to \quad
M_t = Z_t - 2N_\downarrow 
\end{equation}
Since 9 minority bands are fully occupied, we obtain the simple ''rule of 18''
for half-metallicity in $C1_b$ Heusler alloys
\begin{equation}
M_t = Z_t - 18
\end{equation}
the importance of which has been recently pointed out by Jung et al. \cite{jung} 
and Galanakis et al. \cite{Galanakis2002a}. It is a direct analogue to the well-known 
Slater-Pauling behavior of the binary transition metal alloys \cite{Kubler84}. 
The difference with respect to these alloys is, that in the half-Heusler 
alloys the minority population is fixed to 9, so that the screening is achieved 
by filling the majority band, while in the transition metal alloys the majority 
band is filled with 5 $d$-states and charge neutrality is achieved by filling the 
minority states. Therefore in the TM alloys the total moment is given by 
$M_t = 10 - Z_t$. Similar rules with integer total moments are also valid 
for other half-metals, e.g. for the full-Heusler alloys like Co$_2$MnGe with 
$L2_1$ structure. For these alloys 
we will in section 3 derive the ``rule of 24'': $M_t = Z_t - 24$, 
with arises from the fact that the minority band contains 12 electrons. For the 
half-metallic zinc-blende compounds like CrAs the rule is: $M_t = Z_t - 8$, since 
the minority As-like valence bands accommodate 4 electrons \cite{GalanakisZB}. In all cases the moments 
are integer. 

\begin{figure}
\centering
\includegraphics[height=7cm]{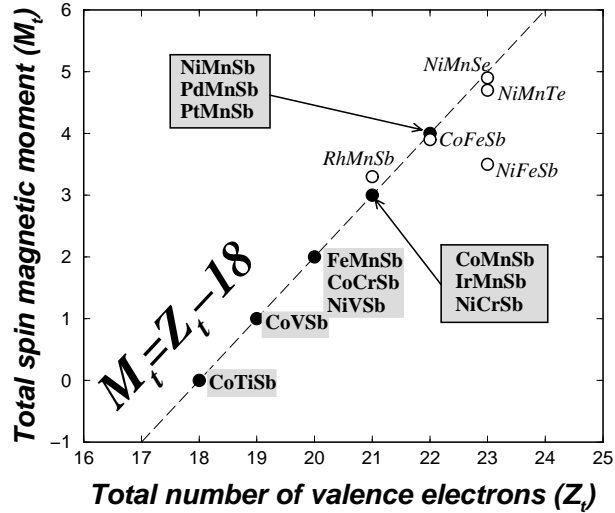}
\caption{Calculated total spin moments for all the studied half
Heusler alloys. The dashed line represents the Slater-Pauling
behavior. With open circles we present the compounds deviating
from the SP curve.}
 \label{figios6}
\end{figure}

In figure \ref{figios6} we have gathered the calculated total
spin magnetic moments for the half-Heusler alloys which 
we have plotted  as a
function of the total number of valence electrons. The dashed
line represents the rule $ M_t = Z_t -18$ obeyed by these
compounds. The total moment  $M_t$ is an integer
quantity, assuming the values 0, 1, 2, 3, 4 and 5 if $Z_t \ge$18.
The value 0 corresponds to the semiconducting phase and the value
5 to the maximal moment when all 10 majority $d$-states are
filled.  Firstly we varied  the valence of the lower-valent
(\textit{i.e.} magnetic) transition metal atom. Thus we substitute
V, Cr and Fe for Mn in the NiMnSb  and CoMnSb compounds using the
experimental lattice constants of the two Mn compounds. For all
these compounds we find that the total spin moment scales
accurately with the total charge and that they all present the
half-metallicity.

As a next test we have substituted Fe for Mn in CoMnSb and NiMnSb,
but both CoFeSb and NiFeSb loose their half-metallic
character. In the case of NiFeSb the majority $d$-states are
already fully occupied as in NiMnSb, thus the additional electron
has to be screened by the minority $d$-states, so that the Fermi
level falls into the minority Fe states and the half-metallicity
is lost; for half-metallicity a total moment of 5 $\mu_B$ would be
required which is clearly not possible. For CoFeSb the situation
is more delicate. This system has 22 valence electrons and if it
would be a half-metal, it should have a total spin-moment of
4 $\mu_B$ as NiMnSb. In reality our calculations indicate that
the Fermi level is slightly above the gap and the total
spin-moment is slightly smaller than 4 $\mu_B$. The Fe atom
possesses a comparable spin-moment in both NiFeSb and CoFeSb
compounds contrary to the behavior of the V, Cr and Mn atoms.
Except NiFeSb other possible compounds with 23 valence electrons
are NiMnTe and NiMnSe. We have calculated their magnetic
properties using the lattice constant of NiMnSb. As shown in
figure \ref{figios6}, NiMnSe almost makes the 5 $\mu_B$ (its
total spin moment is 4.86 $\mu_B$) and 
 is nearly half-metallic, while its isovalent, NiMnTe, has a slightly smaller 
spin moment. NiMnSe and NiMnTe  show
big changes in the majority band compared to systems with 22 
valence electrons as NiMnSb or NiMnAs, since
antibonding $p$-$d$ states, which are usually above $E_F$, are
shifted below the Fermi level, thus increasing the total moment to
 nearly 5 $\mu_B$.

\section{Full Heusler Alloys}
\label{secios:3}

\subsection{Electronic Structure of Co$_2$MnZ with Z = Al, Si, Ga, Ge and Sn}
\label{secios:3-1}

The second family of Heusler alloys, which we discuss, are
the full-Heusler  alloys. We consider in particular compounds containing Co and Mn, as these  are the
full-Heusler alloys that have attracted most of the attention.
They are all strong ferromagnets with high Curie temperatures
(above  600 K) and except the Co$_2$MnAl they show very little
disorder \cite{landolt}. They adopt the $L2_1$ structure shown in 
figure \ref{figios1}. Each Mn or $sp$ atom has eight
Co atoms as first neighbors, sitting in an octahedral symmetry
position, while each Co has four Mn and four $sp$ atoms as first
neighbors and thus the symmetry of the crystal is reduced to the
tetrahedral one.  The Co atoms occupying the two different
sublattices are chemically equivalent as the environment of the
one sublattice is the same as the environment of the second one
but rotated by 90$^o$. The occupancy of two fcc sublattices by Co
(or in general by X) atoms distinguish the full-Heusler alloys
with the $L2_1$ structure from the half-Heusler compounds with the
$C1_b$ structure, like \textit{e.g.} CoMnSb, where only one
sublattice is occupied by Co atoms and the other one is empty.
Although in the $L2_1$ structure, the Co atoms are sitting on
second neighbor positions, their interaction is important to
explain the magnetic properties of these compounds as we will show
in the next section. 

\begin{figure}
\centering
\includegraphics[scale=0.6]{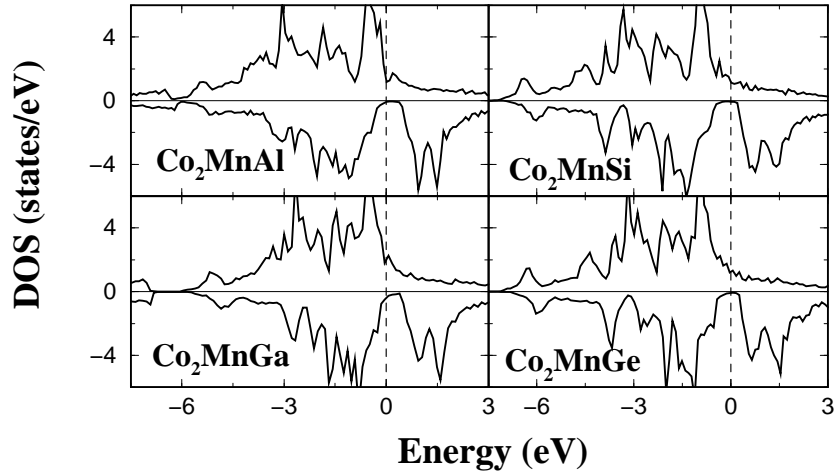}
\caption{Atom-resolved DOS for the Co$_2$MnZ compounds with Z= Al, Si, Ge, Sn
compounds} \label{figios7}
\end{figure}

In figure \ref{figios7} we have gathered the
spin-resolved density of states (DOS) for the Co$_2$MnAl,
Co$_2$MnGa, Co$_2$MnSi and Co$_2$MnGe compounds calculated using
the FSKKR. Firstly as shown by photoemission experiments by Brown
\textit{et al.} in the case of Co$_2$MnSn \cite{Brown98} and
verified by our calculations, the valence band extends around 6  eV below
the Fermi level and the spin-up DOS shows a large peak just below
the Fermi level for these compounds. Although Ishida \textit{et
al.} \cite{Ishida95} predicted them to be half-metals with
small spin-down gaps ranging from 0.1 to 0.3 eV depending on the
material, our previous calculations showed a very small DOS at the Fermi level, 
in agreement with the ASW results of K\"ubler et al. \cite{Kubler84} for Co$_2$MnAl 
and Co$_2$MnSn. However a recalculation of our KKR results with a higher 
$\ell$-cut-off of $\ell_{\rm{max}} = 4$ restores the gap and we obtain good 
agreement with the recent results of Picozzi et al. using the FLAPW method. 
The gap is an indirect gap, with the maximum of the valence 
band at $\Gamma$ and the minimum of the conduction band at the $X$-point. 

In the case of the half-Heusler alloys like NiMnSb the Mn spin
magnetic moment is  very localized due to the exclusion of the
spin-down electrons at the Mn site and amounts to about 3.7
$\mu_B$ in the case of NiMnSb. In the case of CoMnSb the increased
hybridization between the Co and Mn spin-down electrons decreased
the Mn spin moment to about 3.2 $\mu_B$ (in table \ref{tableios1a} we
have gathered the atomic-resolved and total moments of the 
Co$_2$MnZ compounds). In the case of the
full-Heusler alloys each Mn atom has eight Co atoms as first
neighbors instead of four as in CoMnSb and the above hybridization
is very important decreasing even further the Mn spin moment to
less than 3 $\mu_B$ except in the case of Co$_2$MnSn where it is
comparable to the CoMnSb compound.  The Co atoms are
ferromagnetically coupled to the Mn spin moments and they posses a
spin moment that varies from $\sim$0.7 to 1.0 $\mu_B$.
Note that in the half-metallic $C1_b$ Heusler alloys, the $X$-atom has a very 
small moment only, in the case of CoMnSb the Co moment is even negative. 
However in the full Heusler alloys the Co moment is large and positive 
and arises basically from two unoccupied Co bands in the minority conduction 
band, as explained below. Therefore both Co atoms together can have a moment 
of about 2 $\mu_B$, if all majority Co states are occupied. This is basically
the case for Co$_2$MnSi,Co$_2$MnGe and  Co$_2$MnSn (see table 
\ref{tableios1a}). In contrast to 
this the $sp$ atom has a very  
small negative moment which is one order of
magnitude smaller than the Co moment. The negative sign of the
induced $sp$ moment characterizes most of the studied full and
half Heusler alloys with very few exceptions. The compounds
containing Al and Ga have 28 valence electrons and the ones
containing Si, Ge and Sn 29 valence electrons. The first compounds
have a total spin moment of 4$\mu_B$ and the second ones of 5
$\mu_B$ which agree with the experimental deduced moments of these
compounds \cite{Dunlap}. So it seems that the total spin moment,
$M_t$, is related to the total number of valence electrons, $Z_t$,
by the simple relation: $M_t=Z_t-24$, while in the half-Heusler
alloys the total magnetic moment is given by the relation
$M_t=Z_t-18$. In the following section we will analyze the origin
of this rule.
\begin{table}
\centering \caption{Calculated spin magnetic moments in $\mu_B$ using the experimental
lattice constants (see reference \cite{landolt}) for the
Co$_2$MnZ compounds, where Z stands for the $sp$ atom.}
\label{tableios1a}
\begin{tabular}{r|r|r|r|r}\hline\noalign{\smallskip}
$m^{spin}$($\mu_B$) &  Co   &    Mn   & Z  & Total\\ 
\noalign{\smallskip}\hline\noalign{\smallskip} 
Co$_2$MnAl    &  0.768  & 2.530 & -0.096 & 3.970  \\
Co$_2$MnGa    &  0.688  & 2.775 & -0.093 & 4.058 \\
Co$_2$MnSi    &  1.021  & 2.971 & -0.074 & 4.940 \\
Co$_2$MnGe    &  0.981  & 3.040 & -0.061 & 4.941 \\
Co$_2$MnSn    &  0.929  & 3.203 & -0.078 & 4.984 \\
\noalign {\smallskip} \hline
\end{tabular}
\end{table}

\subsection{Origin of the gap in Full-Heusler Alloys}
\label{secios:3-2}

\begin{figure}
\centering
\includegraphics[scale=0.5]{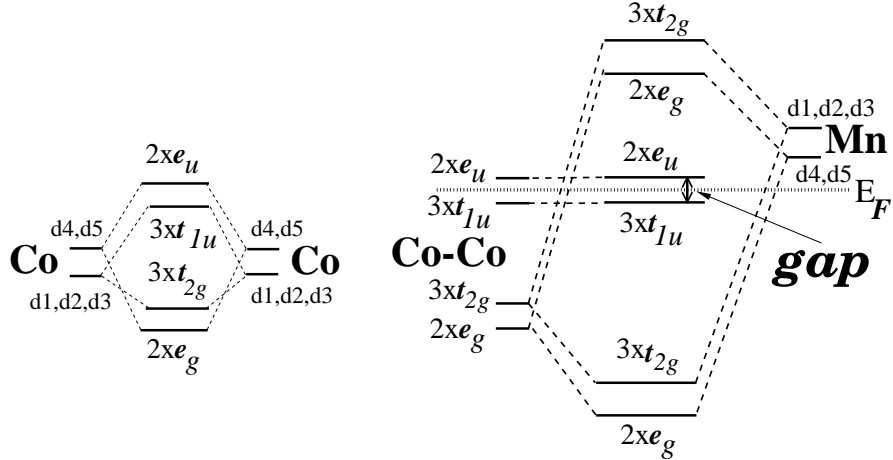}
\caption{Schematic illustration of the origin of the gap in the minority band in 
full-Heusler alloys.} 
\label{figios8}
\end{figure}

Since, similar to the half-Heusler alloys, the 
four $sp$-bands are located far below the Fermi level
and thus are not relevant for the gap,
we consider only the hybridization of the 15 $d$ states of the Mn 
atom and the two Co atoms. For simplicity we consider only the $d$-states at the 
$\Gamma$ point, which show the full structural symmetry. We will give here a 
qualitative picture, since a through group theoretical analysis has been given 
in reference \cite{Galanakis2002b}. Note that the Co atoms form a simple cubic lattice and that the 
Mn atoms (and the Ge atoms) occupy the body centered sites and have 8 Co atoms as nearest neighbors. 
Although the distance between the Co atoms is a second neighbor distance, the 
hybridization between these atoms is qualitatively very important. Therefore we 
start with the hybridization between these Co atoms which is qualitatively sketched 
in figure \ref{figios8}. The 5 $d$-orbitals are divided into the twofold 
degenerate $d_4$, $d_5$ ($r^2$, $x^2 - y^2$)
and the threefold degenerate $d_1$, $d_2$, $d_3$ ($xy, yz, zx$) states. The $e_g$ 
orbitals ($t_{2g}$ orbitals) can only couple with the $e_g$ orbitals $(t_{2g}$ 
orbitals) of the other Co atom forming bonding hybrids, denoted by $e_g$ (or $t_{2g}$) 
and antibonding orbitals, denoted by $e_u$ (or $t_{1u}$). The coefficients in 
front of the orbitals give the degeneracy. 

In a second step we consider the 
hybridization of these Co-Co orbitals with the Mn $d$-orbitals. 
As we show in the second part of figure \ref{figios8}, the
double degenerated $e_g$ orbitals hybridize with the $d_4$ and
$d_5$ of the Mn that transform also with the same representation.
They create a double degenerated bonding $e_g$ state that is very
low in energy and an antibonding one that is unoccupied and above
the Fermi level. The $3\times t_{2g}$ Co orbitals couple to the
$d_{1,2,3}$  of the Mn and create 6 new orbitals, 3 of which are
bonding and are occupied and the other three are antibonding and
high in energy. Finally the $2\times e_u$ and  $3\times t_{1u}$ Co
orbitals \textit{can not} couple with any of the Mn $d$-orbitals since none of these
is transforming with the $u$ representations and they are orthogonal 
to the Co  $e_u$ and  $t_{1u}$ states. With respect
to the Mn and the Ge atoms these states are therefore non-bonding.
The $t_{1u}$ states are below the Fermi level and they are
occupied while the $e_u$ are just above the Fermi level. Thus in
total 8 minority $d$-bands are filled and 7 are empty. 

Therefore all 5 Co-Mn bonding bands are occupied and all 5 Co-Mn antibonding bands 
are empty, and the Fermi level falls in between the 5 non-bonding Co bands, such 
that the three $t_{1u}$ bands are occupied and the two $e_u$ bands are empty. The 
maximal moment of the full Heusler alloys is therefore 7 $\mu_B$ per unit cell, 
which is achieved, if all majority $d$-states are occupied. 

\begin{figure}
\centering
\includegraphics[scale=0.6]{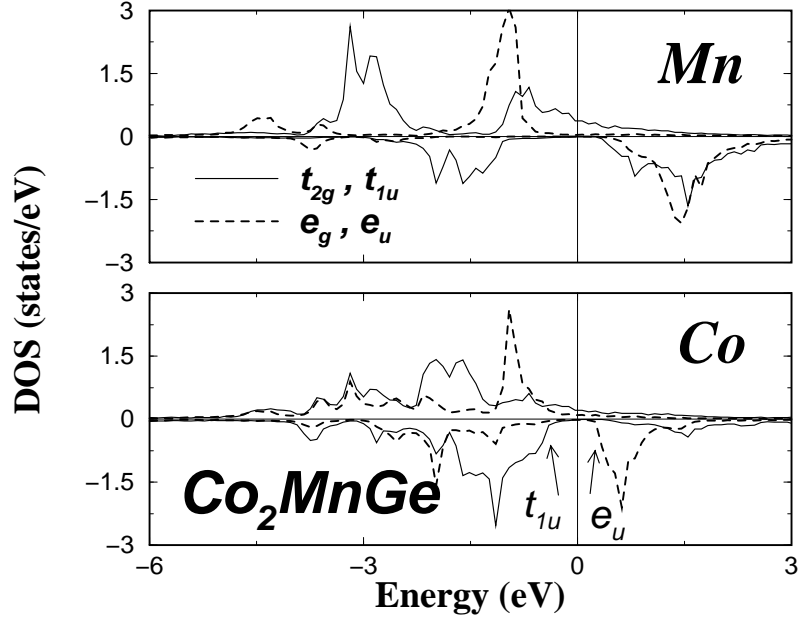}
\caption{Atom- and angular momentum -  resolved DOS for the Co$_2$MnGe 
compound.} \label{figios9}
\end{figure}

In order to demonstrate the existence of the $t_{1u}$ and $e_u$ states at the Fermi 
level, we show in figure \ref{figios9} the LDOS of Co$_2$MnGe 
at the Co and Mn sites, which  are splitted up into the local $d_1, d_2, d_3$ orbitals 
(normally referred to as $t_{2g}$; full lines) and the local $d_4, d_5$ orbitals 
(normally $e_g$; dashed). 
In the nomenclature used above, the $d_1, d_2, d_3$ contributions contain both the 
$t_{2g}$ and the $t_{1u}$ contributions, while the $d_4, d_5$ orbitals contain the 
$e_g$ and $e_u$ contributions. The Mn DOS clearly shows a much bigger effective gap 
at $E_F$, considerably larger than in CoMnSb (figure \ref{figios3}), 
as one would expect from the 
stronger hybridization in Co$_2$MnGe. However the real gap is determined by the Co-Co 
interaction only, in fact by the $t_{1u} - e_u$ splitting, and is smaller than in 
CoMnSb. Thus the origin of the gap in the full-Heusler alloys is rather subtle.

\subsection{Slater-Pauling behavior of the Full-Heusler alloys}
\label{secios:3-3}

\begin{figure}
\centering
\includegraphics[height=7cm]{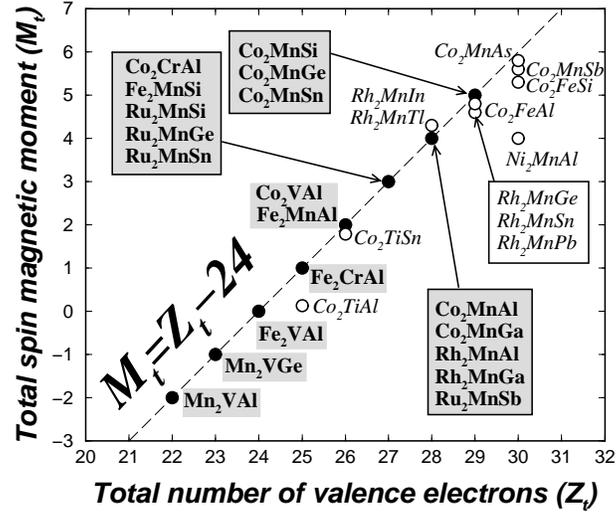}
\caption{Calculated total spin moments for all the studied full
Heusler alloys. The dashed line represents the Slater-Pauling
behavior. With open circles we present the compounds deviating
from the SP curve.}
 \label{figios10}
\end{figure}

Following the above discussion 
we will investigate the Slater-Pauling behavior 
and in figure \ref{figios10} we have
plotted  the total spin magnetic moments for all the compounds
under study as a function of the total number of valence
electrons. The dashed line represents the half metallicity rule: $M_t=Z_t-24$
of the full Heusler alloys. This rule arises from the fact that 
the minority band contains 12 electrons per unit cell: 4 are 
occupying the low lying $s$ and $p$ bands of the $sp$ element
and 8 the Co-like minority $d$ bands ($2\times e_g$, $3\times t_{2g}$ and 
$3\times t_{1u}$), as explained above (see figure \ref{figios8}).
Since 7 minority bands are unoccupied, the largest possible moment is 7 $\mu_B$
and occurs when all majority $d$-states are occupied.

Overall we see that many of our results coincide with the Slater-Pauling
curve. Some of the Rh compounds show small deviations which are
more serious for the Co$_2$TiAl compound. We see that there is no
compound with a total spin moment of 7 $\mu_B$ or even 6 $\mu_B$.
Moreover we found  also examples of half-metallic materials with
less than 24 electrons, Mn$_2$VGe with 23 valence electrons and
Mn$_2$VAl with 22 valence electrons. Firstly, we have calculated
the spin moments of the compounds Co$_2$YAl where Y= Ti, V, Cr, Mn
and Fe. The compounds containing V, Cr and Mn show a similar
behavior. As we substitute Cr for Mn, which has one valence
electron less than Mn, we depopulate one Mn spin-up state and thus
the spin moment of Cr is around 1 $\mu_B$ smaller than the Mn one
while the Co moments are practically the same for both compounds.
Substituting V for Cr has a larger effect since  also the Co
spin-up DOS changes slightly and the Co magnetic moment is
increased by about 0.1 $\mu_B$ compared to the other two compounds
and V possesses a small moment of ~0.2 $\mu_B$. This change in the
behavior is due to the smaller hybridization between the Co atoms
and the V ones as compared to the Cr and Mn atoms. Although all three
Co$_2$VAl, Co$_2$CrAl and Co$_2$MnAl compounds are on the SP curve
as can be seen in figure \ref{figios10}, this is not the case for
the compounds containing Fe and Ti. If the substitution of Fe for
Mn followed the same logic as the one of Cr for Mn then the Fe
moment should be around 3.5 $\mu_B$ which is a very large moment
for the Fe site. Therefore it is energetically more favorable for
the system that also the Co moment is increased, as it was also
the case for the other systems with 29 electrons like Co$_2$MnSi,
but while the latter one makes it to 5 $\mu_B$, Co$_2$FeAl reaches
a value of 4.9 $\mu_B$. In the case of Co$_2$TiAl, it is
energetically more favorable to have a weak ferromagnet than an
integer moment of 1 $\mu_B$ as it is very difficult to magnetize
the Ti atom. Even in the case of the Co$_2$TiSn the calculated
total spin magnetic moment of  1.78 $\mu_B$ (compared to the
experimental value of 1.96  $\mu_B$ \cite{Engen}) arises only from
the Co atoms as was also shown experimentally by Pendl \textit{et
al.} \cite{Pendl}, and the Ti atom is practically nonmagnetic and
the latter compound fails to follow the SP curve.

As a second family of materials we have studied the
compounds containing Fe. Fe$_2$VAl has in total 24 valence
electrons and is a semi-metal, \textit{i.e.} nonmagnetic with a
very small DOS at the Fermi level, as it is already known
experimentally \cite{Fe2VAl}. All the studied Fe compounds follow
the SP behavior as can be seen in figure \ref{figios10}. In the
case of the Fe$_2$CrAl and Fe$_2$MnAl compounds the Cr and Mn
atoms have spin moments comparable to the Co compounds and similar
DOS. In order to follow the SP curve the Fe in  Fe$_2$CrAl is
practically nonmagnetic while in Fe$_2$MnAl it has a small
negative moment. When we substitute Si for Al in Fe$_2$MnAl, the
extra electron exclusively populates Fe spin-up states and the
spin moment of each Fe atom is increased by 0.5 $\mu_B$ contrary
to the corresponding Co compounds where also the Mn spin moment
was considerably increased. Finally we calculated as a test
Mn$_2$VAl and Mn$_2$VGe that  have 22 and 23 valence electrons,
respectively, to see if we can reproduce the SP behavior not only
for compounds with more than 24,
 but also for compounds with less than 24 electrons. As we have already
 shown Fe$_2$VAl is nonmagnetic and Co$_2$VAl, which has two
electrons more, has a spin moment of 2 $\mu_B$. Mn$_2$VAl has two
valence electrons less than Fe$_2$VAl and its total spin moment is
$-\:2\:\mu_B$ and thus it follows the SP behavior; negative total spin moment
means that the ``minority'' band  with the gap has more occupied 
states than the ``majority'' one.

As we have already mentioned the maximal moment of a full-Heusler
alloy is seven $\mu_B$, and should occur, when all 15 majority $d$
states are occupied. Analogously for a half-Heusler alloy the
maximal moment is 5 $\mu_B$. However this limit is difficult to
achieve, since due to the hybridization of the $d$ states with
empty  $sp$-states of the transition metal atoms (sites X and Y in
figure \ref{figios1}), $d$-intensity is transferred into states
high above $E_F$, which are very difficult to occupy. Although in
the case of  half-Heusler alloys, we could identify systems with a
moment of nearly 5 $\mu_B$, the hybridization is much stronger in
the full-Heusler alloys so that a total moment of 7 $\mu_B$ seems
to be impossible. Therefore we restrict our search to possible
systems with 6 $\mu_B$, \textit{i.e.} systems with 30 valence
electrons, but as shown also in figure \ref{figios10}, none of
them makes exactly the 6 $\mu_B$. Co$_2$MnAs shows the largest
spin moment:  5.8 $\mu_B$. 
The basic reason, why moments of 6 $\mu_B$ are so difficult to achieve, 
is that as a result of the strong hybridization with the two Co atoms the 
Mn atom cannot have a much larger moment than 3 $\mu_B$. While due to the 
empty $e_u$-states the two Co atoms have no problem to contribute a total 
of 2 $\mu_B$, the Mn moment is hybridization limited.

\section{Effect of the Lattice Parameter}
\label{secios:4}

\begin{figure}
\centering
\includegraphics[scale=0.6]{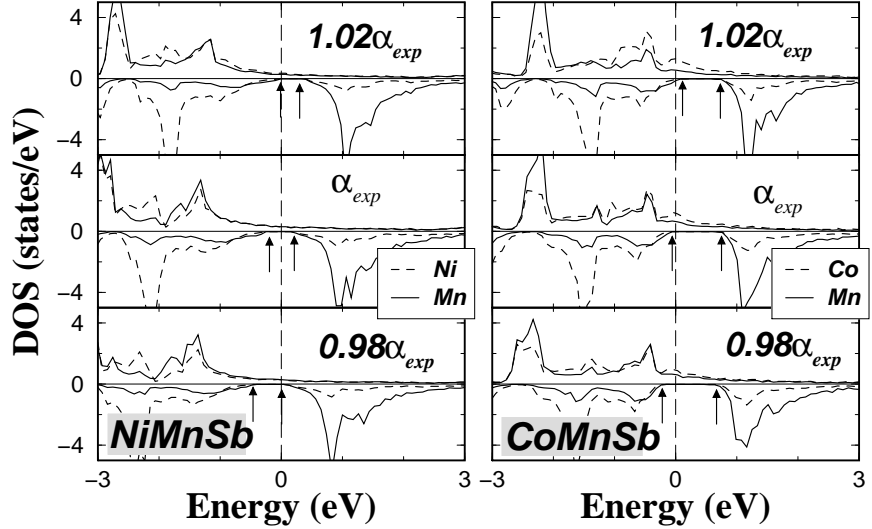}
\caption{Atom-resolved DOS for the experimental lattice parameter for NiMnSb and CoMnSb,
compared with the once compressed or expanded by 2\%. With the small arrow
we denote the edges of the minority gap.} 
\label{figios11}
\end{figure}

In this section we will study the influence of the lattice parameter on the electronic 
and magnetic properties of the $C1_b$ and $L2_1$ Heusler alloys. For these reason we 
plot in figure \ref{figios11} the DOS of NiMnSb and CoMnSb for the experimental lattice 
parameter  and the ones compressed and expanded by 2 \%.
First one sees, that upon compression the Fermi level moves in the direction of the 
conduction band, upon expansion towards the valence band. In both cases, however, 
the half-metallic character is conserved. To explain this behavior, we first note, 
that the Fermi level is determined by the metallic DOS in the majority band. As we 
believe, the shift of $E_F$ is determined from the behavior of the Sb $p$-states, 
in particular by the large extension of these states as compared to the $d$ states. 
Upon compression the $p$-states are squeezed and hybridize stronger, thus pushing 
the $d$-states and the Fermi level to higher energies, i.e. towards the minority 
conduction band. In addition the Mn $d$ and Ni or Co $d$ states hybridize stronger,
which increases the size of the gap. Upon expansion the opposite effects are observed. 
In the case of NiMnSb and for the experimental lattice constant the gap-width
is $\sim$ 0.4 eV. When the lattice is expanded by 2\%\ the gap shrinks to 0.25 eV
and when compressed by 2\%\ the gap-width is increased by 0.1 eV with respect
to the experimental lattice constant and is now 0.5 eV. Similarly in the case
of CoMnSb, the gap is 0.8 eV for the experimental lattice constant, 0.65 for the 
2\%\ expansion and 0.9 eV for the case of the 2\%\ compression.

For the full-Heusler alloys the pressure dependence has been recently studied by 
Picozzi et al. \cite{picozzi} for Co$_2$MnSi, Co$_2$MnGe and Co$_2$MnSn, using both 
the LDA and the somewhat more accurate GGA. 
The general trends are similar: the minority gap increases with compression, 
and the Fermi level moves in the direction of the conduction band. For example
in the case of Co$_2$MnSi the gap-width is 0.81 eV for the theoretical equilibrium
lattice constant of 10.565 \AA . When the lattice constant is compressed to $\sim$
10.15 \AA, the gap-width increases to about 1 eV.

The calculations show that for the considered changes of the lattice constants of 
$\pm$ 2 \%, half-metallicity is preserved. There can be sizeable changes of the local 
moments, but the total moment remains constants, since $E_F$ stays in the gap.

\section{Quaternary Heusler alloys}
\label{secios:5}

We proceed our study by examining the behavior
 of the so-called quaternary Heusler alloys.\cite{GalanakisQuart,ShiraiQuart}. 
In the latter compounds, one of the four
  sites is occupied by two different kinds of  neighboring elements 
like Co$_2$[Cr$_{1-x}$Mn$_{x}$]Al
  where the Y site is occupied by Cr or Mn atoms  To perform this study 
we used the KKR 
method within the coherent potential approximation (CPA) 
as implemented by H. Akai \cite{Akai98}, which 
has been already used with success to study the
magnetic semiconductors \cite{Akai98}. For all calculations we
assumed that the lattice constant varies linearly with the
concentration $x$ which has been verified for several quaternary
alloys \cite{landolt,landolt2}. To our knowledge from the systems
under study only Co$_2$Cr$_{0.6}$Fe$_{0.4}$Al has been studied
experimentally \cite{Felser,Kelekar}.

\begin{figure}
\centering
\includegraphics[scale=0.5]{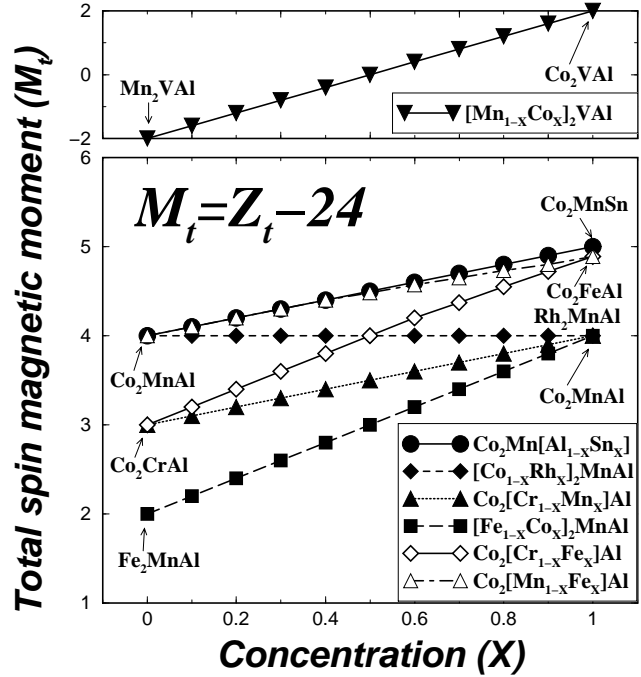}
\caption{Calculated total spin moment $M_t$ in $\mu_B$ for a
variety of compounds as a function of the concentration $x$
($x$=0,0.1,0.2,...,0.9,1). We assumed that the lattice constant
varies linearly with the concentration $x$. With solid lines the
cases obeying the rule $M_t=Z_t-24$ are shown where $Z_t$ and $M_t$ are the average total
number of valence electrons and the average total moment.} \label{figios12}
\end{figure}

We calculated the total spin moment for several quaternary alloys
taking into account several possible combinations of chemical
elements and assuming in all cases a concentration increment of
0.1. We resume our results in figure \ref{figios12}. The first
possible case is when we have two different low-valent transition
metal atoms at the Y site like Co$_2$[Cr$_{1-x}$Mn$_x$]Al. The
total spin moment varies linearly between the 3 $\mu_B$ of
Co$_2$CrAl and the 4 $\mu_B$ of Co$_2$MnAl. In the case of the
Co$_2$[Cr$_{1-x}$Fe$_x$]Al and Co$_2$[Mn$_{1-x}$Fe$_x$]Al
compounds and up to around $x$=0.6 the total spin moment shows the
SP behavior but for larger concentrations it slightly deviates to
account for the non-integer moment value of Co$_2$FeAl 
(see figure \ref{figios11}). This behavior 
is clearly seen in figure \ref{figios12} when we compare the lines for the
Co$_2$[Mn$_{1-x}$Fe$_x$]Al and Co$_2$Mn[Al$_{1-x}$Sn$_x$]
compounds; the latter family follow the SP behavior. The second
case is when one mixes the $sp$ elements, but as we just mentioned
these compounds also obey the rule for the total spin moments. The
third and final case is to mix the higher valent transition metal
atoms like in [Fe$_{1-x}$Co$_x$]$_2$MnAl and
[Rh$_{1-x}$Co$_x$]$_2$MnAl alloys. In the first case the total
spin moment varies linearly between the 2 and 4 $\mu_B$ of
Fe$_2$MnAl and Co$_2$MnAl compounds, respectively. Rh is
isoelectronic to Co and for the second family of compounds we find
a constant integer value of 4 $\mu_B$ for all the concentrations.
A special case is Mn$_2$VAl which has less than 24 electrons and
the total spin moment is -2 $\mu_B$. If now we mix Mn and Co, we
get a family of compounds where the total spin moment varies
linearly between the -2 $\mu_B$ and the 2 $\mu_B$ and  for $x$=0.5
we get the case of a paramagnetic compound consisting of magnetic
elements. Thus all the compounds obey the rule $M_t$=$Z_t$-24,
showing the Slater-Pauling behavior regardless of the origin of
the extra charge.

As a rule of thumb we expect, that for two half-metallic alloys like $XYZ$ and 
$X'YZ$ (or $XY'Z, XYZ'$), which both lay on the Slater-Pauling curve, also the mixtures 
like $X_{1-x}X'_xYZ$ lay on the Slater Pauling curve, with an average moment of 
$<M_t> = (1-x) M^{XYZ}_t + x M_t^{X'YZ}$. However, if these intermediate structures 
are stable, is not guaranteed in particular if the parent compounds are not
neighbors on the Slater-Pauling curve.. 

\section{Point Defects in Half-Metals}
\label{secios:6}

First we would like to discuss some simple rules for point defects in half-metals. 
An important problem is, how and by which states the  additional or missing charge of
 the impurity or point defect is screened. There are several mechanisms:

\begin{enumerate}
\item Either the point defects is screened metallically by the majority states 
such that the number of the minority states does not change.
If $\Delta Z$ is the valence difference of the impurity, then the total change 
$\Delta M_t$ of the alloy moment is given by $\Delta M_t = \Delta Z_t$. 
\item or the point defect is screened by minority states. These can either 
be additional states in the gap, which are introduced by the impurity and 
which, when occupied, lead to $\Delta M_t = - \Delta Z_t$, or these can be 
localized states split-off from the minority band which lead to $\Delta 
M_t = + \Delta Z_t$, if these states are unoccupied. 
\item or both effects occur simultaneously, which is 
not expected for simple defects. 
\end{enumerate}

Note that an isolated point defect cannot change the band gap nor the Fermi level, 
since these are bulk properties. Also the number of minority states cannot be 
changed, except when the defect introduces additional resonances in the minority 
band, or takes out weight from this band by splitting-off states into the gap. As 
a result, in the dilute limit the band gap and half-metallicity is preserved, 
but localized states in the gap, either occupied or empty ones, can occur. An exception
occurs if a multifold degenerate gap-state is partially occupied and thus fixed at 
the Fermi level. Then a symmetry lowering Jahn-Teller splitting of the level is
expected to occur.

For finite concentrations, impurity states in the gap overlap and fast broaden to 
form impurity bands. If the impurities are randomly distributed, one can show by 
applying the coherent potential approximation (CPA) \cite{Akai98}, that the band 
width scales as $\sqrt c$, where $c$ is the impurity concentration. For instance, 
this means, that the bandwidth is for a concentration of 1 \% only a factor 3 
smaller than for 10 \%. Therefore the impurity bands broaden very fast with 
concentration and can soon fill up the band gap, in particular, if the band 
gap is small and the impurity states are rather extended. Therefore the 
control of defects and disorder is an important problem for the application of 
Heusler alloys in spin electronics. 

Unfortunately there are very few theoretical investigations for defects in 
half-metallic Heusler alloys. Picozzi and coworkers have recently studies 
antisite defects in Co$_2$MnGe and Co$_2$MnSi \cite{picozziCo}. For details we 
refer to their review in this volume. Here we address only two screening aspects. 
A Mn-antisite atom on the Co position represents an impurity with $\Delta Z_t = -2$. 
It can either be screened by minority states by pulling two states out-off the 
minority valence band in the energy region above $E_F$. 
However, this is very difficult, since the minority states are basically Co states 
with small Mn admixture. Therefore the Mn antisite is screened by shifting 2 
Mn-states out-off the occupied majority band above the Fermi level, thus 
decreasing the total moment, such that $\Delta M_t = -2$. A Co antisite atom on a 
Mn position is another interesting case with $\Delta Z_t = +2$. Since the majority 
band is nearly filled and a further filling would lead to an increase of the 
moment, which a Co atom cannot sustain, the additional charge is provided by 
the minority states by pulling a double degenerate $e_u$-state from the empty 
Co $e_u$-band in the energy region slightly below $E_F$, thus decreasing the 
moment by $-2 \mu_B$.

Orgassa and coworkers \cite{Orgassa99} have investigated the effect of disorder on the 
electronic structure of NiMnSb. In particular they discuss the effect of impurity 
gap states on the minority DOS and how the impurity bands broaden and fill the gap 
for higher concentrations of antisites.
 While the Mn antisite on Ni position is screened by majority states, 
not leading to a state in 
the gap, the Ni antisite on the Mn position introduces a (three-fold degenerate) 
$d$-level below the Fermi level, which broadens into an impurity band with 
increasing concentration. Already for 1 \% of antisite pairs this band comes 
close to $E_F$ and at 5 \% half-metallicity  disappears and the spin polarization 
at $E_F$ decreases from 100 \% to 52 \%. Thus the general impression is that 
already a disorder concentration of 1 \% is dangerous for the gap and that point 
defects represent a serious problem for half-metallicity. More theoretical studies 
are needed. 

\section{Effect of Spin-orbit Coupling}
\label{secios:7}

As discussed above in a real crystal defects will destroy the 
half-metallic band gap since they destroy the perfect crystal 
periodicity and thus the covalent
hybridization leading to the gap. Moreover at finite temperature thermally 
activated spin-flip scattering, \textit{e.g.} spin waves,
will also induce states within the 
gap \cite{Dowben,Dowben2}.  But even in an ideally prepared
single crystal at zero temperature, the spin-orbit coupling will introduce states in the
half-metallic gap of the minority states (for the spin-down electrons),
which are produced by spin-flip scattering of the majority states
(with spin-up direction). The KKR calculations presented up to this point
were obtained using the scalar-relativistic approximation, thus all-relativistic effects
have been taken into account with the exception of the spin-orbit coupling. 
To study the latter  effect  the KKR 
method has been extended to a fully relativistic treatment by solving the Dirac
equation for the cell-centered potentials \cite{Papanikolaou02}. Thus
the spin-orbit coupling, which is a relativistic effect, is
automatically taken into account. For more details we refer to
reference \cite{Mavropoulos2004}.

Although in our method the Dirac equation is solved, it is easier to
understand the spin-orbit effect within perturbation theory using the
Schr\"odinger equation. In this framework, we remind that the
spin-orbit coupling of the two spin channels is related to the
unperturbed potential $V(r)$ around each atom via the angular momentum
operator $\vec{L}$ and the Pauli spin matrix $\vec{\sigma}$:
\begin{equation}
V_{\mathrm{so}}(r)=\frac{1}{2m^2c^2}\frac{\hbar}{2}
\frac{1}{r}\frac{dV}{dr}\,\vec{L}\cdot\vec{\sigma}
= \left( 
\begin{tabular}{cc}
$V_{\mathrm{so}}^{\uparrow\uparrow}$   & $V_{\mathrm{so}}^{\uparrow\downarrow}$ \\
$V_{\mathrm{so}}^{\downarrow\uparrow}$ & $V_{\mathrm{so}}^{\downarrow\downarrow}$
\end{tabular}
\right)
\label{eq:1.0}
\end{equation}
The $2\times 2$ matrix form is understood in spinor basis.  With
$\uparrow$ and $\downarrow$, the two spin directions are denoted. The
unperturbed crystal Hamiltonian eigenvalues for the two spin
directions are $E^{0\uparrow}_{n\vec{k}}$ and
$E^{0\downarrow}_{n\vec{k}}$, and the unperturbed Bloch eigenfunctions
as $\Psi_{n\vec{k}}^{0\uparrow}$ and
$\Psi_{n\vec{k}}^{0\downarrow}$. Then, noting that within the energy range of the
spin-down gap there exist no unperturbed solutions
$\Psi_{n\vec{k}}^{0\downarrow}$ and $E^{0\downarrow}_{n\vec{k}}$, the
first order solution of the Schr\"odinger equation for the perturbed
wavefunction $\Psi_{n\vec{k}}^{\downarrow}$ reads for states in the gap:
\begin{equation}
\Psi_{n\vec{k}}^{(1)\downarrow}(\vec{r})
=
\sum_{n'} 
\frac{\langle\Psi_{n'\vec{k}}^{0\downarrow}|
V_{\mathrm{so}}^{\downarrow\uparrow}|\Psi_{n\vec{k}}^{0\uparrow}\rangle}
{E_{n\vec{k}}^{0\uparrow}-E_{n'\vec{k}}^{0\downarrow}}
\Psi_{n'\vec{k}}^{0\downarrow}(\vec{r}).
\label{eq:4.0}
\end{equation}
Here, the summation runs only over the band index $n'$ and not over
the Bloch vectors $\vec{k}'$, because Bloch functions with
$\vec{k}'\neq\vec{k}$ are mutually orthogonal. Close to the crossing
point $E_{n\vec{k}}^{0\uparrow}=E_{n'\vec{k}}^{0\downarrow}$ the
denominator becomes small and the bands strongly couple. Then one
should also consider higher orders in the perturbation
expansion. Since at the gap edges there exist spin-down bands of the
unperturbed Hamiltonian, this effect can become important near the gap
edges. Apart from that, the important result is that in the gap region
the spin-down spectral intensity is a weak image of the spin-up
one, determined by the majority states $E_{n\vec{k}}^{0\uparrow}$ in the energy region 
of the gap. Since the spin-down DOS is related to
$|\Psi_{n\vec{k}}^{(1)\downarrow}|^2$, it is expected that within the
gap the DOS has a quadratic dependence on the spin-orbit coupling
strength: $n_{\downarrow}(E) \sim
(V_{\mathrm{so}}^{\downarrow\uparrow})^2$.

\begin{table}
\begin{center}
\caption{Calculated spin polarization at the Fermi level [$P(E_F)$]
  and in the middle of the spin-down gap [$P(E_M)$], for various Heusler
  alloys. The alloys PdMnSb and PtMnSb present a spin-down gap, but
  are not half-metallic, as $E_F$ is slightly below the gap. \label{tableios:2}}
\begin{tabular}{rccccc}
\hline
Compound & FeMnSb & CoMnSb & NiMnSb & PdMnSb & PtMnSb \\ \hline
$P(E_F)$ & 99.3\% & 99.0\% & 99.3\% & 40.0\% &  66.5\% \\
$P(E_M)$ & 99.4\% & 99.5\% & 99.3\% & 98.5\% &  94.5\%  \\ \hline
\end{tabular}
\end{center}
\end{table}

In table \ref{tableios:2} we present the results for several Heusler alloys.
In addition to NiMnSb,  the cases of FeMnSb, CoMnSb,
PdMnSb, and PtMnSb have been studied. Although the last two are not half-metallic 
(a spin-down gap exists, but $E_F$ enters slightly into the valence
band), it is instructive to examine the DOS in the gap region and see
how the spinpolarization decreases as one changes to heavier elements
(Ni$\rightarrow$Pd$\rightarrow$Pt). Here the spinpolarization is $P(E)$ 
is defined by the ratio
\begin{equation}
P(E)=\frac{n^\uparrow (E)-n^\downarrow (E)}{n^\uparrow (E)+n^\downarrow (E)}
\end{equation}
For the Heusler alloys CoMnSb, FeMnSb, NiMnSb,
PdMnSb, and PtMnSb, two quantities are shown: the spin polarization
at $E_F$, $P(E_F)$, and in the middle of the gap, $P(E_M)$.
 The last quantity
reflects the strength of the spin-orbit induced spin flip scattering,
while the first is relevant to our considerations only when $E_F$ is
well within the gap (which is not the case for PdMnSb and
PtMnSb). Clearly, the compounds including 3d transition elements
(NiMnSb, CoMnSb, and FeMnSb) show high spin polarization 
 with small variation as we move along
the 3d row of the periodic table from Ni to Fe. In contrast, when we
substitute Ni with its isoelectronic Pd the number of the induced minority states 
at the  middle of the gap increases and  $P(E_M)$ drops
drastically, and much more when we change Pd for the isoelectronic
Pt. This trend is expected, since it is known that heavier elements
are characterized by stronger spin-orbit coupling.
 But in all cases the alloys retain a region of
very high spin-polarization, instead of a real gap present in the 
scalar relativistic calculations, and thus this phenomenon will not 
be important for realistic applications.
Defects, thermally activated spin-flip scattering, surface or interfaces states
will have a more important effect on the spin-polarization.

Since we can include the spin-orbit coupling to the calculations, 
we are also able to determine the 
orbital moments of the alloys (see reference \cite{GalanakisOrbital} for more details).
The minority valence bands are completely occupied while for the majority ones 
the density of states at the Fermi level is usually very small since most of the
$d$ states are occupied. Thus 
the orbital moments are almost completely quenched and their absolute values 
are negligible with respect to the spin magnetic moments. Also Picozzi and 
collaborators have studied the orbital magnetism of Co$_2$Mn-Si, -Ge and 
-Sn compounds and have reached to similar conclusions \cite{picozzi}.
This also means that the magnetic anisotropy of the half-metallic Heusler alloys
is expected to be rather small.

\section{Surface Properties}
\label{secios:8}

Surfaces can change the bulk properties severely, since the coordination of the 
surface atoms is strongly reduced. Based on the experiences from ferromagnets and semiconductors,
two effects should be particularly relevant for the surfaces of half-metals:
(i) for ferromagnets the moments of the surface atoms are strongly enhanced due to the 
missing hybridization with the cut-off neighbors, and (ii) for semiconductors surface states 
appear in the gap, such that the surface often becomes metallic. Also this is a 
consequence of the reduced hybridization, leading to dangling bond states in the gap.

In this section we review some recent calculations for the (001) surfaces of half- and full-Heusler
alloys. We will not touch the more complex problem of the interfaces with semiconductors, since 
this is treated in the review chapter of Picozzi \textit{et al.} in this book. As a model we
firstly consider the (001)-surface of NiMnSb. Jenkins and King have investigated the MnSb-terminated
NiMnSb(001) surface in detail and have shown that the surface relaxations are very small;
Sb atoms relax outwards while Mn atoms relax inwards with a total buckling of 0.06\AA$\:$ 
\cite{Jenkins01}. Therefore we neglect the relaxations and assume an ``ideal'' epitaxy. The second
possible termination, with a Ni-atom and an unoccupied site at the surface, is not
considered since the configuration
is likely to be unstable or  to show large relaxations of the Ni-surface atoms.

As mentioned above, in the case of the MnSb-terminated (001) NiMnSb surface, Jenkins 
and King have shown that  there are  two surface states \cite{Jenkins01}. The lower lying state 
(at 0.20 eV above the $\bar{\Gamma}$  point) is due to the interaction between 
$e_g$-like dangling bond states 
located at the Mn atoms. The second surface state, which is also higher in energy (0.44 eV above the
$\bar{\Gamma}$  point) rises from the hybridization between $t_{2g}$-like orbitals of Mn with  $p$-type
orbitals of Sb. The first surface state disperses downwards along the [110] direction 
while the second surface state disperses upwards along the same direction. Their behavior is 
inversed along the [1$\bar{1}$0] direction. The two surface states cross along the 
[1$\bar{1}$0] direction bridging the minority gap between the valence and the conduction band.
Along the other directions anticrossing occurs leading to band-gaps. 
Of interest are also the saddle-like structures around the zone center which manifest as van 
Hove singularities in the DOS.

\begin{figure}
\centering
\includegraphics[scale=0.6]{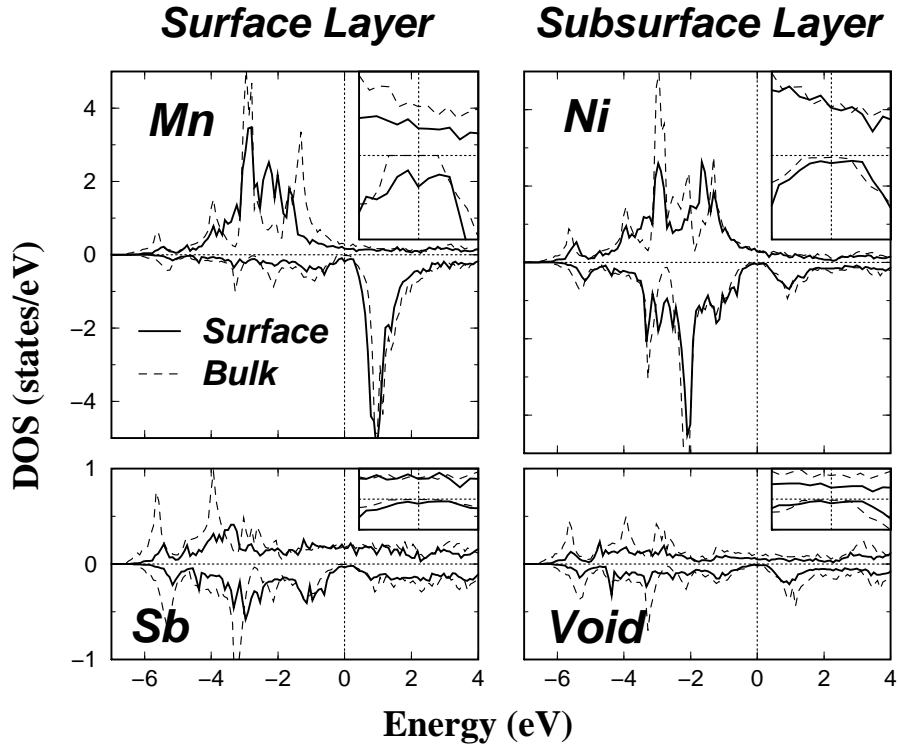}
\caption{Spin- and
atom-projected DOS for the MnSb-terminated NiMnSb(001) surface.
In the insets we have blown up the region around the gap (between -0.5 and 0.5 eV). 
The dashed lines give the local DOS of the atoms in the
bulk.} \label{figios13}
\end{figure}

In figure \ref{figios13}, 
we  present the atom- and spin-projected density
of states (DOS) for the Mn and Sb atoms in the surface layer and
the Ni and vacant site in the subsurface layer for the MnSb
terminated NiMnSb(001) surface. We compare the surface DOS with the bulk 
calculations (dashed line). 
With the exception of the gap region, the surface DOS  is very similar to the bulk
case. The Ni atom in the subsurface layer presents
practically a half-metallic behavior with an almost zero
spin-down DOS, while for the bulk there is an absolute gap. The
spin-down band of the vacant site also presents a very small DOS
around the Fermi level. The Mn and Sb atoms in the surface layer
show  more pronounced differences with respect to the bulk, and
within the gap  there is a very small Mn-$d$ and Sb-$p$ DOS. These 
intensities are due to the two surface states found by Jenkins and King 
\cite{Jenkins01}. These
two surface states are strongly localized  at the surface layer as at the
subsurface layer there is practically no states inside the gap.
Our results are in agreement with the
experiments of Ristoiu \textit{et al.} \cite{Ristoiu00} who  in
the case of a MnSb well ordered (001) surface measured a  high
spin-polarization. Finally we should mention that the spin moment of the Mn atom
at the surface is increased by 0.3 $\mu_B$ with respect to the bulk 
NiMnSb and reaches the $\sim$ 4 $\mu_B$ due to the missing hybridization
with the cut-off Ni neighbors.

It is also interesting to examine the spin-polarization at the
Fermi level. In table \ref{tableios:3} we
have gathered the number of spin-up and spin-down states at the
Fermi level for each atom at the surface and the subsurface layer
for the MnSb-terminated surfaces for different compounds. 
We calculated the spin-polarization as the
ratio between the number of spin-up states minus the number of
spin-down states over the total DOS at the Fermi level. $P_1$
corresponds to the spin-polarization when we take into account
only the surface layer and $P_2$ if we sum the DOS of the surface
and subsurface layers. $P_2$ simulates reasonably well the experimental
situation as the spin-polarization in the case of films is usually
measured by inverse photoemission which probes only the surface of
the sample \cite{Borca02}. In all cases the inclusion of the
subsurface layer increased the spin-polarization since naturally the 
second layer is expected to be more bulk-like. In the case of
the Ni terminated surface, the spin-up DOS at the Fermi level is
equal to the spin-down DOS and the net local polarization $P_2$ is zero.
In the case of the MnSb terminated surface the spin-polarization
increases and now $P_2$ reaches a value of 38\%, which means that
the spin-up DOS at the Fermi level is about two times the
spin-down DOS. The main
difference between the two different terminations is the
contribution of the Ni spin-down states. In the case of the MnSb
surface the Ni in the subsurface layer has a spin-down DOS at the
Fermi level of 0.05 states/eV, while in the case of the
Ni-terminated surface the Ni spin-down DOS at the Fermi level is
0.40 states/eV decreasing considerably the spin-polarization for
the Ni terminated surface; the Ni spin-up DOS is the same for both
terminations. It is interesting also to note that for both
surfaces the net Mn spin-polarization is close  to zero while Sb
atoms in both cases show a large spin-polarization and the number
of the Sb spin-up states is similar to the number of Mn spin-up
states, so that Sb and not Mn is responsible for the large
spin-polarization of the MnSb layer in both surface terminations.
The calculated $P_2$ value of 38\% for the MnSb terminated surface
is smaller than the experimental value of 67\% obtained by Ristoiu
and collaborators \cite{Ristoiu00} for a thin-film terminated in a
MnSb stoichiometric alloy surface layer. But experimentally no
exact details of the structure of the film are known and the
comparison between experiment and theory is not straightforward.

\begin{table}
\begin{center}
\caption{Atomic-resolved spin-up and spin-down local DOS at the Fermi
level in states/eV units.  Polarization ratios at the Fermi level are
calculated taking into account only the surface layer $P_1$, and
the sum of the surface and subsurface layers $P_{2}$. \label{tableios:3}}
\begin{tabular}{r|c|c|c|c|r|r} \hline
& \multicolumn{4}{c|}{MnSb (MnGe or CrAl) -termination}& & \\ \hline &
\multicolumn{2}{c|}{Surface Layer} &\multicolumn{2}{c|}{Subsurface
Layer}&&\\ & Mn ($\uparrow/ \downarrow$) & Sb ($\uparrow/
\downarrow$) & Ni[Co,Pt] ($\uparrow / \downarrow$) & Void
($\uparrow / \downarrow$) & $P_1$  ($\uparrow - \downarrow \over
\uparrow + \downarrow$) & $P_{2}$ ($\uparrow - \downarrow \over
\uparrow + \downarrow$)
\\ NiMnSb & 0.16/0.19 &0.17/0.03 & 0.28/0.05 & 0.05/0.02& 26\% &
38\% \\ CoMnSb & 0.23/0.27 &0.16/0.07 & 0.91/0.15 & 0.07/0.02 &
6\% & 46\% \\ PtMnSb & 0.21/0.24 &0.31/0.06 & 0.38/0.04 &
0.08/0.02 & 26\% & 46\% \\ \hline
\end{tabular}
\end{center}
\end{table}

\begin{figure}
\centering
\includegraphics[scale=0.6]{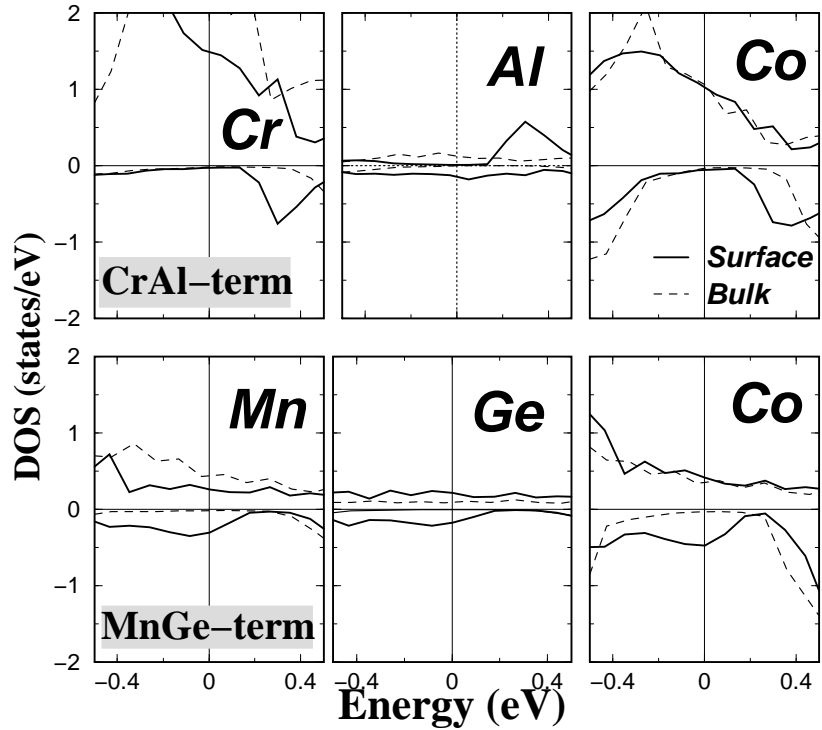}
\caption{Atom- and
spin-projected DOS for the MnGe and CrAl terminated (001) surfaces 
of Co$_2$MnGe and Co$_2$CrAl, respectively. 
With dashed line: the bulk results.} \label{figios14}
\end{figure}

In the case of the full-Heusler alloys containing Mn 
the results are similar to the ones obtained for the NiMnSb compound. 
For the MnGe-terminated (001) surface the induced Mn minority surface states
locally completely kill the spin polarization. This is clearly seen in the
bottom panel of figure \ref{figios14} where we have plotted the DOS 
around the Fermi level for the Mn and Ge atoms at the surface layer and the Co
atoms at the subsurface layer. Note that also for Co at the subsurface the 
local net spin polarization is zero. 
But in the case of Co$_2$CrAl results differ considerably. In line with
the reduction of the total valence electrons by 2, the Cr moment
is rather small (1.54 $\mu_B$) yielding a total moment of only 3
$\mu_B$ instead of 5 $\mu_B$ for Co$_2$MnGe. The Co terminated
Co$_2$CrAl(001) surface shows a similar behavior as the
corresponding surface of Co$_2$MnGe, being in both cases
dominated by a strong Co peak in the gap region of the minority
band. However the CrAl terminated Co$_2$CrAl surface behaves very
differently, being driven by the large surface enhancement of the
Cr moment from 1.54 $\mu_B$ to 3.12 $\mu_B$. As a consequence the
splitting of the Cr peaks in the majority and minority bands is
even enlarged and in particular in the minority band the pseudogap
is preserved. Thus this surface is a rare case, since for all the
other surfaces studied in this paper, the half-metallicity is
destroyed by surface states. If we look closer at the gap region 
(see figure \ref{figios14}) for the CrAl surface we find 
that the Al DOS still has some weight in the gap region. Thus 
compared to NiMnSb there is still one surface state left, which 
has only an Al $p$-component, but no $d$-admixture from the Cr atom.
However in total the surface keeps a high degree of spin polarization:
\textit{i.e.}  $P_2$ is  84\%.

The increase of the spin magnetic moment of the surface atoms due to their 
lower coordination with respect to the bulk can be better understood in the 
case of the transition-metal pnictides and chalcogenides which crystallize
in the zinc-blende structure. Using CrAs and CrSe as examples, Galanakis and 
Mavropoulos have shown that the Cr-terminated (001) surfaces of these alloys retain 
the half-metallicity of the bulk compounds \cite{GalanakisZB}. Contrary to the CrAl-terminated
Co$_2$CrAl(001) there is no $sp$ atom at the surface layer now to induce a surface
state. Thus the situation is simpler to understand as compared to the Heusler alloys.
In the bulk case, Cr has four As or Se atoms as first neighbors. Thus the bonding
 can be described in terms
of four directional bonds around each Cr or As(Se) atom, similar to the case of binary semiconductors 
where $sp^3$ hybrids are being formed. Each Cr atom  provides 0.75 $e^-$ per Cr-As
bond and 0.5 $e^-$ per Cr-Se bond. The Cr atom at the surface looses two out of its four As or Se 
first neighbors and thus regains 1.5 electrons in the case of CrAs(001) and 1 electron in the case of CrSe(001).
The extra electrons now fill up only spin up states and thus the 
total spin moment at the surface (adding the spin moments of the Cr surface atom and of the As or Se 
at the subsurface layer)  is enhanced by exactly 1.5 $\mu_B$ and 1 $\mu_B$ with respect to the 
bulk for CrAs and CrSe, respectively \cite{GalanakisZB}. This rule for the total spin moment 
can be also generalized to the case of the interfaces with semiconductors with the 
condition that  half-metallicity is preserved \cite{GalanakisZB}.

\section{Summary and Outlook}
\label{secios:9}

In this review we have given an introduction into the electronic structure 
and the resulting magnetic properties of half-metallic Heusler alloys, which represent
interesting hybrids between metallic ferromagnets and 
semiconductors. Many unusual features arise from the half-metallicity
induced by the gap in the minority band, and therefore the understanding of the gap 
is of central importance.

For the half-Heusler alloys like NiMnSb, crystallizing in the 
$C1_b$ structure, the gap arises from the hybridization between the 
$d$-wavefunctions of the lower-valent transition metal atom (\textit{e.g.}
Mn) with the $d$-wavefunctions of the higher-valent transition metal
atom (\textit{e.g.} Ni). Thus the $d$-$d$ hybridization leads to 5 
occupied bonding bands, which have a larger Ni and smaller Mn admixture.
These states form the valence band, being separated by a band gap 
from the conduction band which is  formed by five antibonding hybrids with a
large Mn $d$- and a  small Ni $d$-admixture. The role of the $sp$ atoms like
Sb is very different. Firstly they are  important for the bonding, in particular
for the stabilization of the $C1_b$ structure. Secondly the $sp$ atom 
 creates for each spin direction
one $s$ and three $p$ bands in the energy region below the $d$ states which by hybridization
can  accommodate also transition metal electrons, such that \textit{e.g.} Sb
formally acts like a Sb$^{-3}$ and Sn as a Sn$^{-4}$ anion. In this way the effective number
of valence $d$-electrons can be changed by the valence of the $sp$ elements.

Since the minority valence band consist of 9 bands, compounds with 18 valence electrons
like CoTiSb have the same density of states for both spin directions and are semiconductors.
More general, compounds with a total number of $Z_t$ valence electrons per
unit cell are ferromagnets and have an integer total spin moment of $M_t=Z_t-18$,
since $Z_t-18$ is the number of uncompensated spins. For instance, NiMnSb has 22 valence
electrons and therefore a total moment of exactly 4 $\mu_B$.
This relation is similar to the well known Slater-Pauling behavior observed
for binary transition-metal alloys and allows to classify the half-metallic
$C1_b$ Heusler alloys into classes with integer moments between 0 and 5 $\mu_B$.
The maximum moment of 5 $\mu_B$ is difficult to achieve, since it requires that all
majority $d$-sates are occupied.

In the case of the full-Heusler alloys like Co$_2$MnGe, there are,  in addition 
to the Co-Mn bonding
and antibonding $d$-hybrids, also Co states which cannot hybridize with both  the Mn
and the Ge atoms and are exclusively localized at the two Co sublattices. Thus in 
addition to the 5 Co-Mn 
bonding and 5 Co-Mn antibonding bands, there exist 5 such ``non-bonding'' bands 
which are only splitted-up
by the weaker Co-Co hybridization into 3 occupied $d$ states of $t_{1u}$ symmetry and 
2 unoccupied $e_u$ states,
which are located just below and just above the Fermi level such that 
the indirect gap in these materials is smaller than in the half-Heuslers. 
Due to the additional 3 occupied $t_{1u}$ cobalt bands, 
the full-Heusler alloys have 12 occupied minority bands instead of 9 
in the case of the half-Heusler compounds and their relation for 
the total spin magnetic moment becomes $M_t=Z_t-24$. Thus systems like Fe$_2$VAl
with 24 valence electrons are semiconductors, Co$_2$VAl (26 
valence electrons) has  a total spin moment of 2 $\mu_B$, Co$_2$CrAl 3 $\mu_B$,
Co$_2$MnAl 4 $\mu_B$  and finally Co$_2$MnSi which has 29 valence electrons 
has  a total spin moment of 5 $\mu_B$. 
The maximal total spin moment for these alloys is 7  $\mu_B$, but as has been shown
even the 6  $\mu_B$ are unlikely to be achieved.

Having understood the basic elements of the electronic structure, there is still 
a long way to go for understanding the half-metallic behavior of the real materials.
Since the existence of the minority gap is central for any application of
half-metals in spintronics, it is of great importance to understand and control all mechanisms
that can destroy the gap. Firstly we have discussed that spin-orbit interaction
couples the two spin-bands and induces states in the gap; however this effect is weak and 
the spinpolarization remains in most cases as high as $\sim$99 \%. 
While this effect exists already in the ground state, there are, secondly,
excitation effects leading to states in the gap. In the simplest approach     
one can consider in the adiabatic approximation ``static spin waves'', which are 
superpositions of spin-up and spin-down states. At a finite temperature the spinwaves excitations
will then smear out the gap \cite{Dowben2}. A more ambitious treatment
of the interaction with magnons leads to non-quasiparticle excitations in the minority gap
above the Fermi level \cite{Chioncel} and at the Fermi level the minority density of
states strongly depends on the temperature.

Thirdly and most importantly all kind of defects are expected to lead to states in the 
gap. We have discussed this qualitatively for the case of point defects arising from substitutional
disorder and refer to the contribution of Picozzi \textit{et al.} in this volume for
realistic calculations. We feel that many more calculations and experiments are needed; 
the aim is to find systems, which either do not lead to states in the gap (which
is presumably not possible) or systems with particularly high defect formation energies
or sufficiently low annealing temperatures. Equally important is the control of
surface and interface states in the gap, the latter are    in particular important for interfaces to semiconductors.
Here it should be possible to find junctions which do not have
interface states at the Fermi level. The case of Co$_2$CrAl discussed in this paper 
shows, that the occurrence of the transition-metal induced surface state at the 
Fermi level can be suppressed by the increase of the Cr spin moment at the surface.
Finally we add that from  point of view of transport a single interface state does not affect
the magnetoconductance since the wavefunction is orthogonal to all bulk states
incident to the interface. It is the interaction with other defect states in bulk systems and/or
with surface defects which make these states conducting.


\end{document}